\documentclass[twocolumn,showpacs,superscriptaddress,preprintnumbers,prd]{revtex4-1}
\usepackage{graphicx,amsmath,amssymb,dcolumn,bm}
\usepackage{fixmath}
\usepackage{feynmp}
\usepackage{epic,eepic}
\usepackage{slashed}
\usepackage{color}

\definecolor{mycolor}{rgb}{0.6,0.0,0.4}

\begin{document}
\preprint{KEK-TH-2111, J-PARC-TH-0163}
\title{Gluon transversity in polarized proton-deuteron Drell-Yan process}
\author{S. Kumano}
\email[]{shunzo.kumano@kek.jp}
\affiliation{KEK Theory Center,
             Institute of Particle and Nuclear Studies, \\
             High Energy Accelerator Research Organization (KEK), \\
             Oho 1-1, Tsukuba, Ibaraki, 305-0801, Japan}
\affiliation{J-PARC Branch, KEK Theory Center,
             Institute of Particle and Nuclear Studies, KEK, \\
           and Theory Group, Particle and Nuclear Physics Division, 
           J-PARC Center, \\
           Shirakata 203-1, Tokai, Ibaraki, 319-1106, Japan}
\affiliation{Department of Particle and Nuclear Physics, \\
             Graduate University for Advanced Studies (SOKENDAI), \\
             Oho 1-1, Tsukuba, Ibaraki, 305-0801, Japan} 
\author{Qin-Tao Song}
\email[]{qintao@post.kek.jp}
\affiliation{KEK Theory Center,
             Institute of Particle and Nuclear Studies, \\
             High Energy Accelerator Research Organization (KEK), \\
             Oho 1-1, Tsukuba, Ibaraki, 305-0801, Japan}
\affiliation{Department of Particle and Nuclear Physics, \\
             Graduate University for Advanced Studies (SOKENDAI), \\
             Oho 1-1, Tsukuba, Ibaraki, 305-0801, Japan}
\affiliation{School of Physics and Microelectronics, Zhengzhou University, \\
             Zhengzhou, Henan 450001, China}
\date{February 28, 2020}

\begin{abstract}
Nucleon spin structure functions have been investigated mainly 
by longitudinally-polarized ones for finding the origin of 
the nucleon spin. Other types of spin structure functions are 
transversely-polarized ones. In particular, quark transversity 
distributions in the nucleons have very different properties 
from the longitudinally-polarized quark distribution functions,
especially in scaling violation, because they are decoupled from 
the gluon transversity, due to the fact that they are 
helicity-flip (chiral-odd) distributions.
Such studies are valuable for finding not only the origin 
of the nucleon spin but also a signature on physics 
beyond the standard model, because the electric dipole moment 
of the neutron is proportional to the transversity distributions.
Now, there is experimental progress on the quark transversity 
distributions; however, there is no experimental information 
on gluon transversity. In fact, the gluon transversity does not 
exist for the spin-1/2 nucleon due to the helicity-conservation constraint.
One needs a hadron with spin more than or equal to one, 
so that the helicity flip of two units is allowed.
A stable spin-1 target is, for example, the deuteron for studying
the gluon transversity. In this work, we propose a possibility 
for finding the gluon transversity at hadron-accelerator facilities,
especially in the proton-deuteron Drell-Yan process
with the linearly-polarized deuteron, by showing theoretical formalism 
and numerical results. In the experiment, the information 
on the angular distribution of the dimuon is necessary
in the final state; however, the proton beam does not have to be polarized.
We show the dependencies of the Drell-Yan cross section
on the dimuon-mass squared $M_{\mu\mu}^{\,2}$, 
the dimuon transverse-momentum $q_T$,
the dimuon rapidity $y$ in the center-of-momentum frame, 
and the magnitude of the gluon transversity $\Delta_T g$.
We also show typical spin asymmetries in the Drell-Yan process.
Since the internal spin-1/2 nucleons within the deuteron 
cannot contribute directly to the gluon transversity, 
it could be a good observable to find a new non-nucleonic 
component beyond the simple bound system of nucleons in nuclei.
\end{abstract}
\maketitle

\section{Introduction}
\label{intro}

Although the nucleon spin is one of fundamental physics quantities,
its origin is not understood yet. It used to be interpreted by
a combination of three spin-1/2 quarks according to the basic 
quark model \cite{quark-model}. Namely, if two quark spins are 
aligned to the nucleon spin and the other quark spin is opposite,
the nucleon spin should be understood. This simple description
had been taken as granted for a long time until
the European Muon Collaboration (EMC) experiment 
found that this picture is basically wrong in 1988 
by showing that the contribution from the quark spin
is a small fraction \cite{emc-1988}.
We now know that gluon-spin and partonic orbital-angular-momentum
contributions could be significant as sources of the nucleon spin.

Since the EMC discovery, studies on high-energy polarized-hadron reactions 
have been done to clarify the origin of the nucleon spin, mainly through 
longitudinally-polarized structure functions \cite{nucleon-spin}. 
In addition, there had been discussions how to decompose the nucleon spin 
into quark-gluon spin components and orbital-angular-momentum contributions
in a gauge invariant way \cite{decomposition}.
There are also studies in lattice QCD 
on the spin decomposition \cite{lattice-deomposition}. 
Furthermore, efforts have been made recently to obtain $x$-dependent 
parton distributions from lattice QCD \cite{lattice-pdfs}. 
Now, experimental clarifications on gluon-spin and partonic 
orbital-angular-momentum contributions become necessary. 
For probing the orbital-angular-momentum part,
we need to investigate three-dimensional structure functions
\cite{gpds-gdas}, namely, 
generalized parton distributions (GPDs) \cite{gpds},
generalized distribution amplitudes (GDAs) \cite{gdas}, and
transverse-momentum-dependent parton distributions (TMDs) \cite{tmds}.
Such experimental studies are under investigations
at experimental facilities in the world \cite{sk-dis2018}.

In spite of much progress on longitudinal spin physics,
the transversely-polarized structure functions are not known well
\cite{ralston-soper,artru-mekhfi,jaffe-ji,Artru-2002,transversity-pr-2002,br-book},
although there are some recent studies on quark transversity distributions
\cite{transversity-pdfs}. 
Such studies provide an important and alternative information
in solving the nucleon spin puzzle.
In particular, the quark transversity distributions of the nucleon
are decoupled from the gluon transversity in the $Q^2$ evolution
\cite{transversity-q2,transversity-q2-gluon}
due to the helicity-flip (chiral-odd) property, 
which is an important difference from the longitudinal spin.
Therefore, studies of the transversity distributions are
other tests of our understanding on nucleon spin by different observables.
In addition, since electric dipole moments of hadrons,
such as the neutron, are proportional to the transversity
distributions \cite{dipole-m}, the transversity studies 
are valuable also for searching physics beyond the standard model
by measuring the electric dipole moments.
There are also transversity GPD studies \cite{trans-gpds}.

For understanding of transverse-polarization physics, the gluon transversity
distribution should be investigated in addition to the quark transversity.
The gluon transversity is not experimentally measured at this stage,
whereas we have a rough idea on the quark transversity distributions
\cite{transversity-pdfs}. However, there are future experimental
projects to measure them accurately 
at Thomas Jefferson National Accelerator Facility (JLab) and
Electron-Ion Collider (EIC) \cite{future-q-transversity,eic-2016}.
Therefore, much progress is expected for the gluon transversity
in the near future because of the JLab experiment on the gluon transversity 
with the polarized-deuteron target \cite{jlab-gluon-trans}.

On the other hand, independent experiments are desirable
at other experimental facilities, especially 
at hadron accelerator facilities, to probe different 
kinematical regions of the gluon transversity from the JLab one.
In particular, the large $Q^2$ region 
($M_{J/\psi}^{\, 2} < Q^2 < M_{\Upsilon}^{\, 2}$)
should be measured by the Drell-Yan process,
in comparison with the JLab one
typically from a few GeV$^2$ to several GeV$^2$.
The Fermilab spin-physics project SpinQuest is under preparation
as the E1039 experiment \cite{Fermilab-dy}, 
and the proton-deuteron Drell-Yan will be also possible 
with the polarized deuteron target.

The purpose of this work is to propose a new process to measure
the gluon transversity for the first time at hadron facilities.
Especially, we propose that the gluon transversity
should be measured in the proton-deuteron Drell-Yan process
by considering the Fermilab-E1039 experimental project. 
However, our formalism can be used in principle for 
the Drell-Yan experiments at any high-energy hadron
accelerator facilities.

We may remind the reader that a hadron with spin at least one
is necessary for studying the gluon transversity, which does not
exist for the spin-1/2 nucleon, because the change of two units of spin 
($\Delta s = 2$) is necessary for the gluon transversity \cite{br-book}. 
The purpose of our work is to propose 
a possible process to probe the gluon transversity 
at hadron accelerator facilities
as an alternative and independent method 
from the lepton scattering measurement at JLab and EIC.

Apart from the spin physics, the gluon transversity is
a theoretically interesting quantity to probe an exotic aspect 
of the deuteron, hadrons, and nuclei. 
For example, the deuteron is a weak-bound state of a proton
and a neutron mainly in the S wave with a small probability
of D-state admixture. Since the internal nucleons do not contribute
directly to the gluon transversity due to the spin-1/2 nature,
the gluon transversity of the deuteron is expected 
to be a small quantity. However, if a finite distribution is
found in future, it could indicate an existence of a non-nucleonic 
component or some other exotic hadronic mechanism within the deuteron. 
There are theoretical-model \cite{transversity-model}
and lattice-QCD \cite{transversity-lattice} studies on this topic.
For example, a contribution from the $\Delta\Delta$ component
in the deuteron was estimated in Ref.\,\cite{transversity-model}
as a possible finite gluon transversity in the deuteron.
Therefore, the gluon transversity distributions are interesting
quantities, which shed light on unknown exotic aspects 
in the deuteron and nuclei beyond the simple bound systems 
of protons and neutrons.

In addition to the transversity, there are related studies 
on polarized deuteron structure functions.
For example, the tensor-polarized structure function $b_1$ will
be measured in the near future at JLab \cite{fs83,hjm-89,jlab-b1},
and polarized proton-deuteron reactions \cite{pd-drell-yan}
could be investigated at Fermilab for measuring 
tensor-polarized distribution functions \cite{Fermilab-dy,fermilab-pd}.
Since the conventional deuteron model cannot explain existing
experimental measurements by the HERMES collaboration \cite{b1-convolution},
a new hadronic mechanism would be needed for their interpretation
\cite{miller-b1}.
Such new aspects of the deuteron at high energies may be related
to the gluon transversity distributions because 
they probe non-nucleonic component in the deuteron.

In this article, the transversity distributions are explained 
by starting from the basic Pauli-Lubanski operator 
and matrix-element forms of local quark current operators
in comparison with the longitudinally-polarized 
parton distribution functions (PDFs) in Sec.\,\ref{transversity}.
Since the quark-transversity distributions are 
directly related to electric dipole moments of hadrons, 
the relation is briefly explained.
Then, the gluon transversity is explained. It exists
in hadrons only with spin larger than or equal to one 
due to the helicity conservation. 
Next, the theoretical formalism is shown
for the proton\,($p$)-deuteron\,($d$) Drell-Yan process $p + d \to \mu^+\mu^+ +X$
in Sec.\,\ref{formalism}, including kinematical variables,
polarizations of spin-1 deuteron, hadron correlation functions
in terms of the PDFs, and expression of
the cross section $p + d \to \mu^+\mu^+ +X$.
Then, partonic matrix elements and actual cross sections are obtained.
The proton-deuteron Drell-Yan cross sections are numerically
shown in Sec.\,\ref{results}, and 
our studies are summarized in Sec.\,\ref{summary}.

\vspace{-0.15cm}
\section{Transversity distributions}
\label{transversity}

\vspace{-0.15cm}
\subsection{Pauli-Lubanski operator in Poincar\'e group \\
and quark transversity distributions}
\label{Pauli-Lubanski}
\vspace{-0.10cm}

The transversity is not a popular terminology outside 
the high-energy spin-physics community, and the transverse 
spin and polarization are somewhat confusing as shown in 
Eqs.\,(\ref{eqn:nucleon-polarization-trans-1}) and 
(\ref{eqn:nucleon-polarization-trans-2}), 
so that its basics are first explained \cite{br-book}. 

The four-dimensional space-time coordinate transformation
${x'}^\mu=\Lambda^\mu_{\ \nu} x^\nu +a^\mu$ is called 
the inhomogeneous Lorentz transformation or the Poincar\'e transformation.
The invariance under this transformation is a fundamental symmetry 
in quantum field theory. Representations of the Poincar\'e group
are classified by Casimir operators $p^2$ and $W^2$
\cite{Itzeykson-Zuber}.
Here, $p^\mu$ is the momentum operator which is the generator of translations, 
and $W^\mu$ is the Pauli-Lubanski operator which is 
the generator of Lorentz transformations.
The Pauli-Lubanski operator is defined by the angular-momentum operator
$J^{\nu\rho}$ and the momentum as
\begin{align}
W_\mu = \frac{1}{2} \varepsilon_{\mu\nu\rho\sigma} J^{\nu\rho} p^\sigma,
\label{eqn:Pauli-Lubanski}
\end{align}
with the antisymmetric tensor definition $\varepsilon_{0123}=+1$.
The angular-momentum operator is given by
\begin{align}
J^{\mu\nu} = \frac{1}{2} \sigma^{\mu\nu} 
     + \left ( x^\mu p^\nu - x^\nu p^\mu \right ) .
\label{eqn:angular-momentum}
\end{align}
Here, the antisymmetric tensor $\sigma^{\mu\nu}$ is defined by
$\sigma^{\mu\nu} = \frac{i}{2} (\gamma^\mu \gamma^\nu -\gamma^\nu \gamma^\mu )$.
The eigenvalues of $p^2$ and $W^2$ are $M_N^{\,2}$ and $-M_N^{\,2} s (s+1)$, 
respectively:
\begin{align}
\! \! \! 
p^2 \left | p\, s \rangle \right.  
   = M_N^{\,2} \left | p\, s \rangle \right. \! ,  \ \ 
W^2 \left | p\, s \rangle \right.  
   = -M_N^{\,2} s (s+1) \left | p\, s \rangle \right. \! ,
\label{eqn:nucleon-states-pauli-lub}
\end{align}
where $M_N$ and $s(=1/2)$ are mass and spin of the nucleon.

From the Pauli-Lubanski operator, the polarization operator $\Pi$
could be defined for the nucleon as \cite{br-book}
\begin{align}
\! \! \! \! 
\Pi \equiv -\frac{1}{M_N} W \cdot s \,
= \frac{1}{2 M_N}  \gamma_5 \, \slashed{s} \, \slashed{p} \,
= \frac{1}{2 M_N  i}  \gamma_5 
\sigma_{\mu\nu} \,
s^{\mu}  p^{\nu} ,
\label{eqn:polariztion-Sigma}
\end{align}
where it is expressed by the spin and momentum vectors.
The spin vector $s^\mu$ satisfies $s^2 = -1$ and $s \cdot p=0$, 
and it is generally expressed as
\begin{align}
s^\mu = \left ( \,   
        \frac{\vec p \cdot \hat n}{M_N},\, 
         \hat n + \frac{\vec p \cdot \hat n}{M_N \, (M_N+p^0)}  \vec p
        \, \right ) ,
\label{eqn:nucleon-spin}
\end{align}
where $\hat n$ is a unit vector in three-dimensional space
to indicate the spin-polarization direction.
The longitudinal polarization is given by 
$\hat n = \pm \vec p / | \vec p |$, and the transverse one
is by $\hat n = \hat n_\perp$ where $\hat n_\perp$
is a two-dimensional transverse unit vector. 
The polarization vector of Eq.\,(\ref{eqn:polariztion-Sigma}) 
becomes the helicity operator
\begin{align}
\Pi_{\parallel} 
= \frac{1}{2} \frac{\vec \Sigma \cdot \vec p}{|\, \vec p \, |} 
                = \frac{1}{2} \Sigma_{\parallel}
                = \frac{1}{2} \gamma_5 \, \gamma_0 \gamma_{\parallel} 
                = \frac{\sigma_{\parallel}}{2}
\left(
    \begin{array}{cc}
      I &  0  \\
      0 &  I  
    \end{array}
\right) , 
\label{eqn:nucleon-polarization-long}
\end{align}
if the nucleon is longitudinally polarized. 
Here, the longitudinal direction is taken along the third axis
($| \, \vec p \, | =p_\parallel = p_3$).
The spin operator $\vec \Sigma$ is defined by 
\begin{align}
\vec \Sigma = \gamma_5 \, \gamma_0 \vec \gamma,
\label{eqn:spin-sigma}
\end{align}
$\sigma_{\parallel}$ is the longitudinal Pauli-spin matrix defined by
$\sigma_{\parallel} \equiv \vec\sigma \cdot \vec p /| \, \vec p \, |$,
and $I$ is the $2 \times 2$ identity matrix.
This helicity operator commutes with the free-quark Hamiltonian 
$H_0 = \alpha_3 \, p_3 = \gamma_0 \, \gamma_3 \, p_3$, 
so that it is a conserved quantity. 

On the other hand, if the nucleon is transversely polarized, 
the transverse polarization and spin operators may be given,
from Eqs.\,(\ref{eqn:polariztion-Sigma}) and (\ref{eqn:spin-sigma}), 
by
\begin{align}
\Pi_\perp = 
\frac{1}{2 M_N} \, \gamma_5 \, \slashed{s}_\perp \, \slashed{p}, \ \ 
\Sigma_{\perp} = \gamma_5 \, \gamma_0 \gamma_{\perp} ,
\label{eqn:nucleon-polarization-trans-1}
\end{align}
where $s_\perp^{\,\mu} = (0,\, \hat n_\perp) \equiv n_\perp^{\,\mu}$,
and $\Sigma_{\perp}$ and $\gamma_{\perp}$ are defined by
$\Sigma_{\perp} = \hat n_\perp \cdot \vec\Sigma $ and
$\gamma_{\perp} = \hat n_\perp \cdot \vec\gamma 
   = \gamma_1 \cos\phi_\perp + \gamma_2 \sin\phi_\perp $
with the azimuthal angle $\phi_\perp$ of $\hat n_\perp$.
However, they do not commute with the free Hamiltonian.
It means that there is no eigenstates of 
$\Pi_\perp$ or $\vec\Sigma_{\perp}$ with $H_0$,
so that the quarks in the transversely-polarized nucleon
cannot have a definite transverse-spin state
with the polarization or spin operator
of Eq.\,(\ref{eqn:nucleon-polarization-trans-1}).
However, if the transverse-polarization operator $\widetilde \Pi_\perp$ 
is defined with an extra $\gamma_0$ with $\Sigma_\perp$ by
\begin{align}
\! \! \! \! 
\widetilde \Pi_\perp
\equiv
 \frac{1}{2} \, \gamma_0 \, \Sigma_\perp
            =  \frac{1}{2} \,  \gamma_5 \, n_\perp \cdot \gamma
            = \frac{\sigma_{\perp}}{2}
\left(
    \begin{array}{cc}
      I &  0  \\
      0 &  -I  
    \end{array}
\right) \! , 
\label{eqn:nucleon-polarization-trans-2}
\end{align}
it commutes with $H_0$. Here, 
the transverse Pauli spin matrix is 
defined by $\sigma_{\perp} = \hat n_\perp \cdot \vec\sigma $.
We call this polarization as ``transversity".
It means that the quarks exist in the nucleon as
a definite transverse-polarization state for $\widetilde \Pi_\perp$, 
although the spin eigenstate does not exist 
for the operators $\Pi_\perp$ and $\vec\Sigma_{\perp}$.
The transversity distributions are denoted $\Delta_T q$ 
for quarks and $\Delta_T g$ for gluon throughout this article.
However, there are other notations $h_T$ \cite{ralston-soper}, 
$\Delta_1 q$ \cite{artru-mekhfi}, $h_1$ \cite{jaffe-ji}, 
and $\delta q$ \cite{Artru-2002} for quarks and other ones 
($\Delta_2 G$, $a$, $\Delta_L g$, $\delta G$, $h_{1TT,g}$, $\Delta_T g$),
as shown later in Eq.\,(\ref{eqn:gluon-transversity-definition}),
for gluon, so that one may pay attention to the notation 
in reading past papers on the transversity.

Since the transverse spin $s_\perp$ was mentioned for the nucleon,
it is briefly explained in the following.
The polarized charged-lepton deep inelastic scattering (DIS) from
a polarized nucleon is described by the antisymmetric hadron tensor
in terms of two polarized structure functions $g_1$ and $g_2$:
\begin{align}
W_{\mu\nu}^{(A)} & =  \frac{2 \, M_N}{p\cdot q} \, \varepsilon_{\mu\nu\alpha\beta}
  \,  q^\alpha \left [  s^\beta g_1 + \left ( s^\beta 
              - \frac{s \cdot q}{p \cdot q} p^\beta \right )  g_2  \right ]
\nonumber \\
& =  \frac{2 \, M_N}{p\cdot q} \, \varepsilon_{\mu\nu\alpha\beta}
  \,  q^\alpha  \left (  s^\beta_\parallel g_1 
            +s^\beta_\perp g_T      \right ) ,
\label{eqn:wmunu-a}
\end{align}
where $g_T$ is defined by 
\begin{align}
g_T \equiv g_1 + g_2 .
\label{eqn:gt-g1-g2}
\end{align}
The nucleon spin vector is decomposed into the longitudinal 
and transverse ones as
\begin{align}
s^\mu = s^\mu_\parallel + s^\mu_\perp
      = \frac{\lambda_N}{M_N} p^\mu + s^\mu_\perp ,
\label{eqn:nucleon-l-t-spin}
\end{align}
where $\lambda_N$ is the nucleon helicity.
It indicates that the transverse-spin $s^\mu_\perp$ contribution
to the cross section is suppressed by the factor $M_N/p^+$,
where $p^+$ is the lightcone momentum 
$p^+ = (p^0 + p^3)/\sqrt{2}$,
in comparison with the longitudinal term.
As explained, the transverse spin is not a conserved quantity,
the structure function $g_T$ does not allow a simple probabilistic
interpretation of the leading-twist level, although it can be
measured experimentally in the standard polarized charged-lepton
scattering measurement.

Therefore, the transverse-polarization physics is investigated 
by the transversity distributions in the twist-2-level collinear framework.
Experimentally, the quark transversity distributions are determined, 
for example, by analyzing the data of semi-inclusive DIS process 
and proton-proton collisions with dihadron production \cite{transversity-pdfs}.
There are other possibilities such as semi-inclusive hadron-production
processes and Drell-Yan processes for the quark transversity distribution.
However, there is little information on how to determine the gluon transversity
at this stage. Especially, there is no article to investigate the gluon
transversity by using hadron experimental facilities as far as we are aware.
Here, a possible process is proposed in this work 

\subsection{Longitudinally-polarized \\ and transversity distributions for quarks}
\label{q-transversity}

\begin{figure}[b]
 \vspace{-0.00cm}
\begin{center}
   \includegraphics[width=4.2cm]{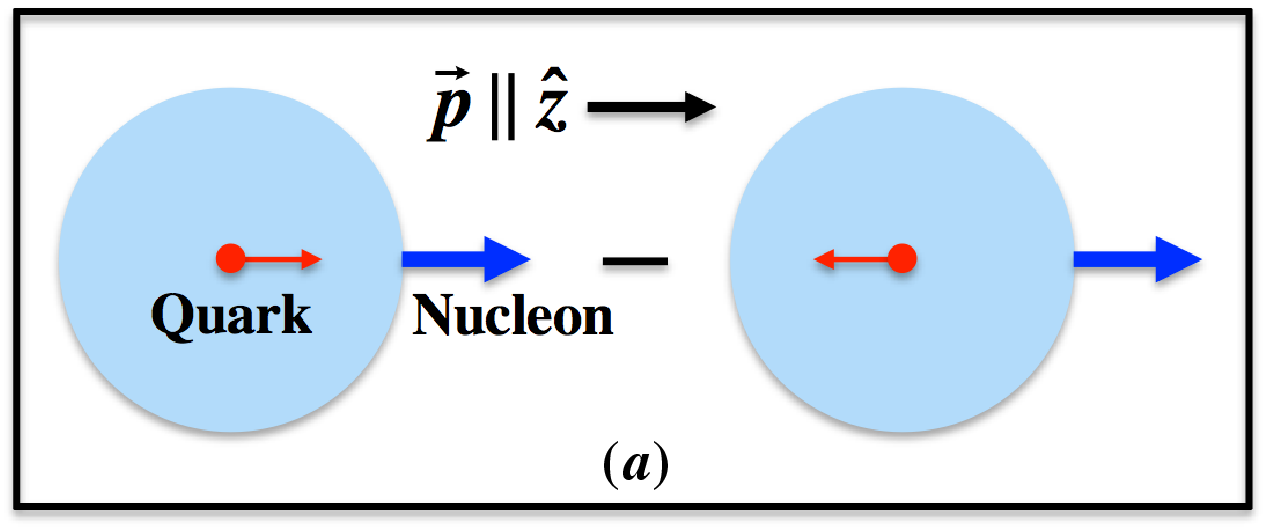}
   \ \ \ 
   \includegraphics[width=3.7cm]{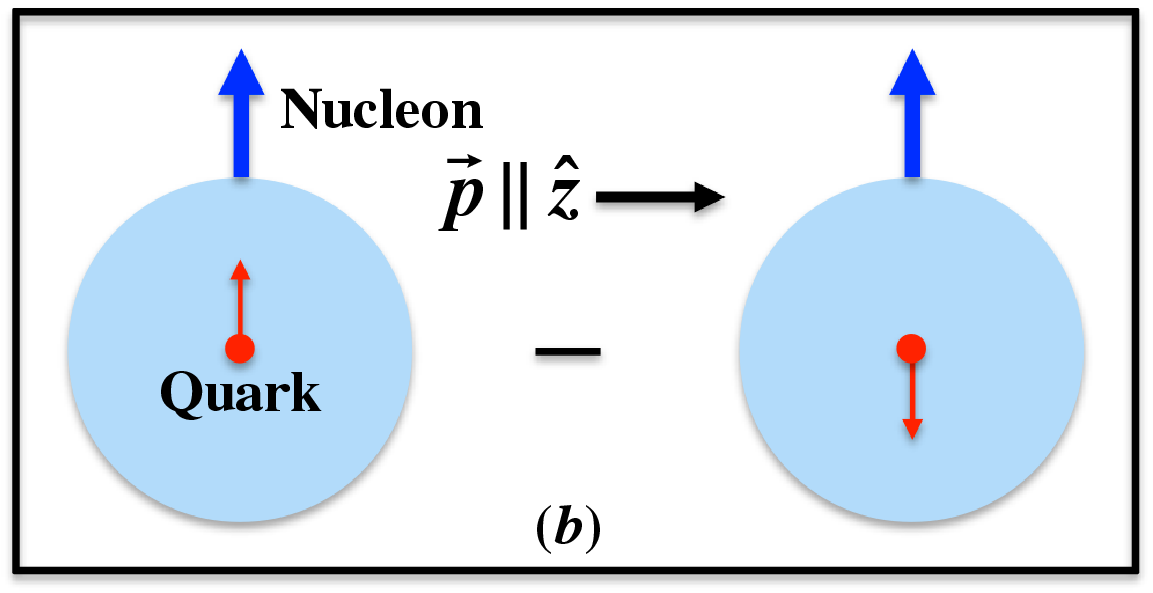}
\end{center}
\vspace{-0.5cm}
\caption{Illustration of $(a)$ longitudinally-polarized quark distribution
and $(b)$ quark transversity distribution.}
\label{fig:long-trans-spin}
\vspace{-0.30cm}
\end{figure}

First, we explain quark transversity distributions.
The longitudinally-polarized quark distribution functions are given
by the difference between the quark distributions with spin parallel to
the nucleon spin and the ones with antiparallel spin:
$\Delta q (x) = q_+ (x) - q_- (x)$,
where $x$ is the momentum fraction carried by a quark,
as illustrated in Fig.\,\ref{fig:long-trans-spin}$(a)$.
Here, $+$ and $-$ indicate parallel and
anti-parallel quark spins to the longitudinal nucleon spin.
They are relatively well determined now for the nucleon
by polarized lepton DIS and polarized proton-proton collisions.
For the transversely polarized nucleon, similar distributions
called transversity distributions are expressed as
$\Delta_T q (x) = q_\uparrow (x) - q_\downarrow (x)$,
where $\uparrow$ and $\downarrow$ indicate parallel and
anti-parallel quark polarizations,
as defined by the polarization operator of 
Eq.\,(\ref{eqn:nucleon-polarization-trans-2}),
to the transversely-polarized nucleon spin,
as illustrated in Fig.\,\ref{fig:long-trans-spin}$(b)$.

The unpolarized, longitudinally-polarized, and transversity 
distribution functions are defined for quarks
by the following matrix elements \cite{br-book}:
\begin{align}
& \! \! 
q (x)  = \! \int  \frac{d \xi^-}{4\pi} \, e^{i x p^+ \xi^-}
\langle \, p  \left | \, \bar\psi (0) 
\gamma^+  \psi (\xi)  \, \right | p \, \rangle 
_{\xi^+=\vec\xi_\perp=0}  ,
\nonumber \\[-0.00cm]
& \! \! 
\Delta q (x) = \! \int \! \frac{d \xi^-}{4\pi} \, e^{i x p^+ \xi^-}
\langle \, p \, s_L \! \left | \, \bar\psi (0) 
\gamma^+ \gamma_5 \psi (\xi)  \, \right | p \, s_L \, \rangle 
_{\xi^+=\vec\xi_\perp=0}  ,
\nonumber \\[-0.00cm]
& \! \! 
\Delta_T q (x)  = \! \int  \frac{d \xi^-}{4\pi} \, e^{i x p^+ \xi^-}
\nonumber \\[-0.05cm]
& \ \ \ \ \ 
\times
\bigg\langle \, p \, s_{T j} \left | \, \bar\psi (0)  
\, i \, \gamma_5 \, \sigma^{j +} 
 \psi (\xi) \right | p \, s_{Tj} \, \bigg\rangle 
_{\xi^+=\vec\xi_\perp=0} ,
\label{eqn:delta-deltaT-qx}
\end{align}
where $s_L$ and $s_{Tj}$ ($j=1$ or $2$) indicate longitudinal 
and transverse polarizations of the nucleon,
and $\psi$ is the quark field.
Here, gauge links for satisfying the color gauge invariance
are abbreviated because they do not play an important role
in the collinear PDFs.
These distribution functions are leading twist (twist-2) ones.
In Sec.\,\ref{Pauli-Lubanski}, we introduced the structure function
$g_T$ associated with the transverse spin. 
It is also written in an operator matrix element in the similar way as
\begin{align}
g_{T, q} (x) & =  \frac{p^+}{M_N} \int \frac{d \xi^-}{4\pi} 
\, e^{i x p^+ \xi^-}
\nonumber \\
& \ \ \times
\langle \, p \, s_T \! \left | \, \bar\psi (0) 
\gamma_\perp \gamma_5 \psi (\xi)  \, \right | p \, s_T \, \rangle 
_{\xi^+=\vec\xi_\perp=0}  .
\label{eqn:gt-matrix}
\end{align}
This is a twist-3 structure function, which is suppressed
typically by the kinematical factor of the order of
$M_N/p^+$ in cross sections.
By defining ``good" ($+$) and ``bad" ($-$) components of 
the field $\psi$ as 
$\psi =\psi_+ + \psi_-$, 
$\psi_\pm \equiv (\gamma^\mp \gamma^\pm /2 ) \psi$
\cite{jaffe-ji},
the operator of the longitudinally-polarized quark distributions
$\Delta q (x)$ in Eq.\,(\ref{eqn:delta-deltaT-qx})
is expressed by 
$\bar\psi \gamma^+ \gamma_5 \psi = \bar\psi_+ \gamma^+ \gamma_5 \psi_+$,
namely only by the good components.
For the distribution $g_{T, q} (x)$, it is given only by
the cross combination of $\psi_+$ and $\psi_-$ as
$\bar\psi \gamma_\perp \gamma_5 \psi 
 = \bar\psi_+ \gamma_\perp \gamma_5 \psi_- 
 + \bar\psi_- \gamma_\perp \gamma_5 \psi_+$,
which indicates that it is a twist-3 distribution.

\begin{figure}[b]
 \vspace{-0.00cm}
\begin{center}
   \includegraphics[width=5.0cm]{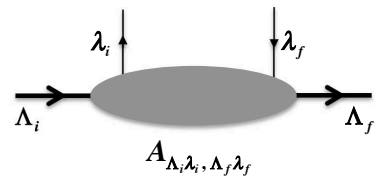}
\end{center}
\vspace{-0.7cm}
\caption{Parton-hadron forward scattering amplitude 
$A_{\Lambda_i \lambda_i,\, \Lambda_f \lambda_f}$
with the hadron helicities
$\Lambda_i$ and $\Lambda_f$ and parton ones $\lambda_i$ and $\lambda_f$.}
\label{fig:helicity-amp}
\vspace{-0.30cm}
\end{figure}

As known in the DIS formalism, 
structure functions of the nucleon are given by the imaginary
part of forward scattering amplitudes by the optical theorem.
The PDFs are expressed by parton-hadron forward scattering amplitudes 
illustrated in Fig.\,\ref{fig:helicity-amp}.
The amplitude is denoted as 
$A_{\Lambda_i \lambda_i ,\, \Lambda_f \lambda_f}$
with the initial and final hadron helicities
$\Lambda_i$ and $\Lambda_f$ and parton ones $\lambda_i$ and $\lambda_f$.
The helicity conservation indicates the relation
\cite{Jaffe-1996}
\begin{align}
\Lambda_i - \lambda_i = \Lambda_f - \lambda_f ,
\label{eqn:heclity-conserv}
\end{align}
where the minus signs exist for the parton helicities 
due to out-going and in-coming particles
in comparison with the hadron helicities.
The PDFs are related to the helicity amplitudes as \cite{br-book,Jaffe-1996}
\begin{align}
\! \! \! \! \! \!
q (x) &          = q_+ (x) + q_- (x) 
                 \sim \text{Im} \, (A_{++,\, ++} + A_{+-,\, +-}) ,
\nonumber \\
\! \! \! \! \! \!
\Delta q (x) &   = q_+ (x) - q_- (x) 
                 \sim \text{Im} \, (A_{++,\, ++} - A_{+-,\, +-}) ,
\nonumber \\
\! \! \! \! \! \!
\Delta_T q (x) & = q_\uparrow (x) - q_\downarrow (x) 
                 \sim \text{Im} \, A_{++,\, --} \ .
\label{eqn:delta-deltaT-qx-amplitudes}
\end{align}
The last relation $\Delta_T q \sim \text{Im} \, A_{++,\, --}$
is given in Ref.\,\cite{Jaffe-1996}.
It should be noted that the helicity amplitudes
$A_{\Lambda_i \lambda_i ,\, \Lambda_f \lambda_f}$ 
are used in this article and Ref.\,\cite{Jaffe-1996}.
However, one needs to be careful about a notation difference
in some other papers, because the amplitudes are often defined 
by spin components along the quantization axis as given in Refs.\,\cite{hjm-89}.
For example, the relation is
$\Delta_T q \sim \text{Im} \, A_{-+,\, +-}$ in the
$A_{hH,\, h'H'-}$ notation \cite{jlab-gluon-trans,transversity-lattice}, 
where $h$ and $H$ ($h'$ and $H'$)
are initial (final) quark and hadron spin components.

If the spin states are defined by the transversity basis
$\left | \, \uparrow \, \rangle \right.$ and 
$\left | \, \downarrow \, \rangle \right.$,
they are expressed by the longitudinally-polarized states as
\cite{br-book}
\begin{align}
\left | \, \uparrow \, \rangle \right. 
& = \frac{1}{\sqrt{2}} 
   \left [ \      
     \left | \, + \, \rangle \right. +  \, \left | \, - \, \rangle \right. 
   \, \right ], 
\nonumber \\
\left | \, \downarrow \, \rangle \right. 
& = \frac{1}{\sqrt{2}} 
   \left [ \      
     \left | \, + \, \rangle \right. -  \, \left | \, - \, \rangle \right. 
   \, \right ] ,
\label{eqn:long-trans-states}
\end{align}
where the direction of the polarization $\uparrow$ is
taken along the $x$ axis.
Therefore, if the amplitudes are defined by the transversely-polarized 
states, the transversity distribution is given by
\begin{align}
\Delta_T q (x) & = q_\uparrow (x) - q_\downarrow (x) 
                 \sim \text{Im} \, 
                 (A_{\uparrow \uparrow,\,\uparrow \uparrow} 
                - A_{\uparrow \downarrow,\,\uparrow \downarrow}) .
\label{eqn:deltaT-qx-trans-amplitudes}
\end{align}
The transversity distributions are important 
leading-twist functions for clarifying 
the internal structure of the nucleon.

\subsection{Electric dipole moment of neutron}
\label{diole-moment}

The transversity is an important physics quantity not only 
for clarifying the nature and origin of the nucleon spin 
but also for finding a signature of beyond the standard model 
by observing electric dipole moments of the neutron and other hadrons. 
Therefore, its connection to the electric dipole moment is briefly
explained.
The neutron electromagnetic current is expressed as
\cite{dipole-m}
\begin{align}
\langle n \left | J_\mu^{em} \right | n \rangle
= \bar u (p') \bigg [ & \gamma_\mu F_1 (q^2) 
  + \frac{\kappa}{2M_N}  i \sigma_{\mu\nu} q^\nu F_2 (q^2)
\nonumber \\
& 
 + \frac{d_n}{2M_N} 
 \gamma_5 \sigma_{\mu\nu} q^\nu  F_3 (q^2) 
 \bigg ] u(p) ,
\label{eqn:neutron-em-current}
\end{align}
by including the time-reversal odd term with 
the form factor $F_3$ in addition to the ordinary parity 
and time-reversal even terms with the form factors $F_1$ and $F_2$.
Here, $\kappa$ is the anomalous magnetic moment,
$F_1$ and $F_2$ are Dirac and Pauli form factors,
and they are related to the electric and magnetic form factors as
$G_E (q^2) = F_1 (q^2) + q^2/(2 M_N^2) \kappa F_2 (q^2)$ and
$G_M (q^2) = F_1 (q^2) + \kappa F_2 (q^2)$.
The initial and final neutron momenta are denoted as $p$ and $p'$,
$q$ is the momentum transfer given by $q=p-p'$, 
and $u(p)$ is the Dirac spinor
for the neutron.
The last $F_3$ term is the 
time-reversal odd one, in combination with the electromagnetic
field $A^\mu$ for the Hamiltonian \cite{dipole-m},
with the factor of the neutron electric dipole moment (EDM) $d_n$
in the unit of $e/(2 M_N)$.
The electric dipole form factor $F_3$ is normalized as
$F_3 (0)=1$ at $q^2 = 0$.
Here, we use the function notation $F_3$ which has been used so far
in EDM studies; however, the function $F_3$ is conventionally used 
for the structure function $F_3$ in neutrino scattering. 
They should not be confused.

On the other hand, the neutron EDM is expressed by 
integrals of the transversity distributions, 
so-called the tensor charge $\Delta_T q$, as
\cite{dipole-m}
\begin{align}
d_n   & =  \sum_q d_q \, \Delta_T q , 
\nonumber \\[-0.10cm]
& \Delta_T q  \equiv \int_0^1 dx \, 
        \left [ \Delta_T q (x) - \Delta_T \bar q (x) \right ] ,
\label{eqn:transversity-edm}
\end{align}
where $d_q$ is the quark EDM.
Namely, the neutron EDM is investigated theoretically
by calculating the quark EDMs
in the standard model or some models beyond the standard model,
and they should be multiplied by the tensor charge
in order to compare with experimental measurements.
Therefore, the studies of transversity distributions have impact
on investigations of physics beyond the standard model
by observing the EDMs of the neutron, other hadrons, and nuclei.

\subsection{Gluon transversity in hadrons with spin $\mathbold{\ge}\,$1}
\label{g-transversity}

As shown in Eq.\,(\ref{eqn:delta-deltaT-qx-amplitudes}),
the transversity distribution $\Delta_T q (x)$ is associated with
the quark spin flip ($\lambda_i=+$, $\lambda_f=-$), 
so that it is a chiral-odd distribution.
The quark transversity exists in the nucleon because
the spin flip $\Delta s = 1$ is possible in the spin-1/2 nucleon,
whereas the gluon transversity $\Delta_T g$ does not exist
in the nucleon because the spin flip $\Delta s = 2$ 
is not possible.
Therefore, the quark transversity distributions evolve 
in the scale $Q^2$ without the corresponding gluon distribution 
in the nucleon \cite{transversity-q2,transversity-q2-gluon}.
This situation is very different from the longitudinally-polarized
PDFs, where the quark and gluon distributions couple with each other
in the $Q^2$ evolution as we usually have in the unpolarized PDFs.
This is an important test of perturbative QCD in high-energy spin physics.

\begin{figure}[t]
\begin{center}
   \includegraphics[width=5.0cm]{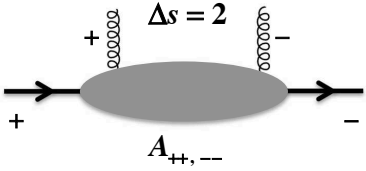}
\end{center}
\vspace{-0.7cm}
\caption{Gluon-hadron forward scattering amplitude 
$A_{++,\, --}$ with the spin flip of 2 ($\Delta s=2$)
for finding the gluon transversity.
The hadron spin should be $s \ge 1$. For example, 
it is spin-1 deuteron.}
\label{fig:helicity-amp-transversity}
\vspace{-0.30cm}
\end{figure}

In the same way with the quark transversity expression of
Eq.\,(\ref{eqn:deltaT-qx-trans-amplitudes}),
the gluon transversity is written by 
the helicity distribution as
\cite{jlab-gluon-trans,transversity-lattice}
\begin{align}
\Delta_T g (x) & \sim \text{Im} \, A_{++,\, --} \ .
\label{eqn:delta-deltaT-gx-amplitudes}
\end{align}
This equation indicates that the spin flip of two units
$\Delta s = 2$ ($|\lambda_f-\lambda_i|=|\Lambda_f-\Lambda_i| = 2$)
is necessary for the gluon transversity $\Delta_T g$, 
and it is illustrated in Fig.\,\ref{fig:helicity-amp-transversity}.
In order to find the gluon transversity, hadrons with spin 
$ \ge 1$ should be used. 
The most simple and stable spin-1 hadron or nucleus
is the deuteron, so that it is used first 
for future experimental studies of the gluon transversity.
There is an experimental proposal to measure it in the polarized 
electron-deuteron DIS by looking at the azimuthal angle of
the deuteron-spin polarization \cite{jlab-gluon-trans}. 
In our work, we investigate a possibility to investigate
the gluon transversity in the deuteron by hadron accelerator facilities
as an alternative way to the lepton-facility measurements.

In the similar way with Eq.\,(\ref{eqn:delta-deltaT-qx}),
the gluon transversity distribution is written 
in the matrix element form 
\begin{align}
\Delta_T g (x) 
& = \varepsilon_{TT,\alpha\beta}
\int  \frac{d \xi^-}{2\pi} \, x p^+ \, e^{i x p^+ \xi^-} 
\nonumber \\[-0.00cm]
& \ \ \ \ 
\times
\langle \, p \, E_{x} \left |  A^{\alpha} (0) \, A^{\beta} (\xi)  
\right | p \, E_{x} \, \rangle 
_{\xi^+=\vec\xi_\perp=0}  ,
\label{eqn:delta-deltaT-gx}
\end{align}
where $\varepsilon_{TT}^{\alpha \beta} = +1$ for $\alpha = \beta =1$,
$\varepsilon_{TT}^{\alpha \beta} = -1$ for $\alpha = \beta =2$,
and the other components are zero, as defined later
in Eq.\,(\ref{eqn:gluon-polarization-tensor-1}).
The notation $E_x$ indicates the linear polarization of the deuteron
along the positive $x$-axis. 
This expression for $\Delta_T g (x)$
is obtained by using the gluon correlation function
of Eq.\,(\ref{eqn:correlation-g-1}) and the collinear expression
of Eq.\,(\ref{eqn:correlation-integrated-g}).
In this equation, $A^\alpha$ is the gluon field 
$A^\alpha =A_a^\alpha t^a$
which includes the SU(3) generator $t^{a}$ expressed by
the Gell-Mann matrix as $t^{a}$ = $\lambda^{a} /2$
with the color index $a$.
Here, the summation is taken over $a$.
Therefore, $\Delta_T g$ is given by the linear polarization difference 
$\varepsilon_x - \varepsilon_y$
for the gluon in the linearly-polarized ($E_x$) deuteron.
The linear polarizations for the deuteron ($E_x$, $E_y$) and the gluon
($\varepsilon_x$, $\varepsilon_y$) are explained 
in Secs.\,\ref{d-polarizations} and \ref{kinematics}.
We should mention that the name ``gluon transversity" is misleading
in the sense that it does not mean the transverse polarization 
of the gluon but it is actually on the linear polarization.

\section{Formalism for Drell-Yan process
         $\mathbold{p + d \to \mu^+\mu^- + X}$}
\label{formalism}

Our formalism is explained for describing the Drel-Yan cross section
$p + d \to \mu^+\mu^- + X$ in this section.
Since this work is on deuteron spin physics, especially on 
gluon transversity, we introduce polarizations of spin-1 deuteron.
Then, the cross section formalism is discussed.

\subsection{Spin-1 deuteron polarizations}
\label{d-polarizations}

Since polarizations of the spin-1 deuteron, which contains
tensor polarizations, are not familiar, they are explained in general 
by using the spin-density matrix.
Let us consider a spin state $| \psi \rangle$ for a particle of spin $s$,
and it is expanded by eigenstates of the $z$-component of 
the spin operator $s_z$ with the expansion coefficients $c_m$ as 
$
\left | \psi \rangle \right. = \sum_m c_m \left | s m \rangle \right. 
$
\cite{leader-book}.
Matrix elements of an operator $\hat O$ are denoted as 
$O_{m'm} = \langle sm' | \hat O | sm \rangle$,
and then expectation value in the state  $| \psi \rangle$ is expressed as
$
\langle \psi | \hat O | \psi \rangle
 = \sum_{m,m'} c_{m'}^* c_m O_{m'm} 
$.
If the state is an incoherent mixture of pure states $| \psi^{(i)} \rangle$ 
with the probability $p^{(i)}$, the expectation value is written as
$
\langle \hat O \rangle = \sum_i p^{(i)} 
\langle \psi^{(i)} | \hat O | \psi^{(i)} \rangle
=
\sum_{m,m'} O_{m'm} \sum_i p^{(i)} c_{m'}^{(i)*} c_m^{(i)} 
$.
Defining the spin-density matrix $\rho_{mm'}$ as
\begin{align}
\rho_{mm'} = \sum_i p^{(i)} c_m^{(i)} c_{m'}^{(i)*} ,
\label{eqn:spin-density}
\end{align}
we obtain
\begin{align}
\langle \hat O \rangle = \sum_{m,m'} O_{m'm} \rho_{mm'}
= \text {Tr} \, (O  \rho).
\label{eqn:expectation}
\end{align}
Therefore, if the spin-density matrix is known, the expectation value
over the ensemble can be calculated for operators.

For example, the density matrix is given by
${\mathbold\rho}_{\scriptscriptstyle{1/2}} 
= (1 +s_i  {\mathbold\sigma}_i )/2$,
where the summation is taken over $i=1$, $2$, and $3$,
with the Pauli matrix ${\mathbold\sigma}_i$,
and the spin-polarization vector 
is given by $s_i = \langle {\mathbold\sigma}_i \rangle = \text {Tr} \, 
({\mathbold\rho}_{\scriptscriptstyle{1/2}} \, {\mathbold\sigma}_i)$.
In the similar way, the density matrix for a spin-1 particle is given
in the Cartesian coordinates as 
\cite{leader-book,bacchetta-2000-PRD,bacchetta-2002-PhD}
\begin{align}
{\mathbold\rho} = \frac{1}{3} \left ( 1 + \frac{3}{2} \,  S_i {\mathbold\Sigma}_i
+ 3 \, T_{ij}  {\mathbold\Sigma}_{ij} \right ) ,
\label{eqn:density-spin-1}
\end{align}
where ${\mathbold\Sigma}_i$ ($i=1, 2, 3$) are $3 \times 3$ spin matrices 
for the spin-1 deuteron:
\begin{align}
{\mathbold\Sigma}_x & = \frac{1}{\sqrt{2}} 
\left(
    \begin{array}{ccc}
      0 &  1 &  0 \\
      1 &  0 &  1 \\
      0 &  1 &  0
    \end{array}
\right) , \ \ 
{\mathbold\Sigma}_y = \frac{i}{\sqrt{2}}  
\left(
    \begin{array}{ccc}
      0 & -1 &  0 \\
      1 &  0 & -1 \\
      0 &  1 &  0
    \end{array}
\right) , 
\nonumber \\
{\mathbold\Sigma}_z & =  
\left(
    \begin{array}{ccc}
      1 &  0 &  0 \\
      0 &  0 &  0 \\
      0 &  0 & -1
    \end{array}
\right) , \ \ 
\label{eqn:spin-1-matrices}
\end{align}
and ${\mathbold\Sigma}_{ij}$ are spin tensors defined by
\begin{align}
{\mathbold\Sigma}_{ij} = 
\frac{1}{2} \left ( {\mathbold\Sigma}_i {\mathbold\Sigma}_j 
                  + {\mathbold\Sigma}_j {\mathbold\Sigma}_i \right ) 
- \frac{2}{3} \, {\mathbold I} \, \delta_{ij}  .
\label{eqn:Tij}
\end{align}
Here, ${\mathbold I}$ is the $3 \times 3$ identity matrix,
and the ${\mathbold\Sigma}_{ij}$ convention
of Refs.\,\cite{bacchetta-2000-PRD,bacchetta-2002-PhD} is used instead of the one 
in Ref.\,\cite{leader-book}.
The spin polarization vector $S_i$ is given by 
\begin{align}
S_i = \langle {\mathbold\Sigma}_i \, \rangle 
= \text {Tr} \, ({\mathbold\rho} {\mathbold\Sigma}_i \, )  ,
\label{eqn:spin-vector-1}
\end{align}
and the tensor $T_{ij}$ is a real and traceless one given by 
\begin{align}
T_{ij} = \langle {\mathbold\Sigma}_{ij} \, \rangle
= \text {Tr} \, ({\mathbold\rho} {\mathbold\Sigma}_{ij}\, )  .
\label{eqn:tensor-1}
\end{align}

The spin vector and tensor are parametrized in the rest frame of
the deuteron as
\cite{bacchetta-2000-PRD,Boer-2016,vonDaal-2016}
\begin{align}
\! \! \! \! 
{\mathbold S} & = (S_{T}^x,\, S_{T}^y,\, S_L) ,
\nonumber \\
\! \! \! \! 
{\mathbold T}  & = \frac{1}{2} 
\left(
    \begin{array}{ccc}
     - \frac{2}{3} S_{LL} + S_{TT}^{xx}    & S_{TT}^{xy}  & S_{LT}^x  \\[+0.20cm]
     S_{TT}^{xy}  & - \frac{2}{3} S_{LL} - S_{TT}^{xx}    & S_{LT}^y  \\[+0.20cm]
     S_{LT}^x     &  S_{LT}^y              & \frac{4}{3} S_{LL}
    \end{array}
\right) .
\label{eqn:spin-1-vector-tensor}
\end{align}
We use the tensor ${\mathbold T}$ in 
Refs.\cite{bacchetta-2000-PRD,Boer-2016,vonDaal-2016},
whereas the factor $-(2/3)S_{LL}$ is denoted as $S_{LL}$ 
in Ref.\,\cite{bacchetta-2002-PhD}.
The spin vector and tensor are written in terms of the polarization
vector $\vec E$ of the deuteron as
\begin{align}
{\vec S} = \text{Im} \, (\, \vec E^{\, *} \times \vec E \,),
\ \ \ 
T_{ij}  = \frac{1}{3} \delta_{ij} 
       - \text{Re} \, (\, E_i^{\, *} E_j \,),
\label{eqn:spin-1-vector-tensor-2}
\end{align}
and their covariant forms are given by
\cite{bacchetta-2002-PhD,hjm-89,jlab-b1}
\begin{align}
S^{\,\mu} & =  \frac{1}{M} \,
\varepsilon^{\,\mu\nu\alpha\beta} \, p_\nu \,
\text{Im} \, (\, E_\alpha^{\, *} E_\beta \,),
\nonumber \\
T^{\,\mu\nu} & = - \frac{1}{3} 
       \left ( g^{\,\mu\nu} - \frac{p^{\,\mu} p^{\,\nu}}{p^2} \right )
       - \text{Re} \, (\, E^{\,\mu \, *} E^{\,\nu} \,) .
\label{eqn:spin-1-vector-tensor-3}
\end{align}
Here, $M$ and $p$ are the deuteron mass and momentum.
At this stage, the deuteron spin quantization axis is taken as
the $z$ direction; however, $-z$ direction is taken later
in calculating the cross section along the deuteron momentum direction.
Then, the deuteron polarization vector $E$
and also gluon polarization vector $\varepsilon$
are defined as
\begin{align}
& E_\pm = \varepsilon_\pm  =\frac{1}{\sqrt{2}} 
       \left ( \, 0,\, \mp 1,\, -i,\, 0 \, \right ) ,  
\ 
E_0 = \varepsilon_0 = \left ( \, 0,\, 0,\, 0,\, 1 \, \right ) ,
\nonumber \\
& E_x = \varepsilon_x  =\frac{1}{\sqrt{2}}
                 \left ( \varepsilon_- - \varepsilon_+ \right )
                 = \left ( \, 0,\, 1,\, 0,\, 0 \, \right ) ,
\nonumber \\
& E_y = \varepsilon_y   =  \frac{i}{\sqrt{2}} 
                    \left ( \varepsilon_- +\varepsilon_+ \right )
                    = \left ( \, 0,\, 0,\, 1,\, 0 \, \right ) . 
\label{eqn:dct}
\end{align}
\begin{widetext}
\noindent
Using these quantities, we express the spin-density matrix of 
Eq.\,(\ref{eqn:density-spin-1}) as
\cite{bacchetta-2000-PRD,Boer-2016,vonDaal-2016}
\begin{align}
{\mathbold\rho}   = 
\left(
    \begin{array}{ccc}
     \frac{1}{3} + \frac{S_L}{2}  + \frac{S_{LL}}{3} 
   & \frac{S_{T}^x - i S_{T}^y}{2 \sqrt{2}} + \frac{S_{LT}^x - i S_{LT}^y}{2 \sqrt{2}} 
   & \frac{S_{TT}^{xx} - i S_{TT}^{xy}}{2}                                  
     \\[+0.20cm]
     \frac{S_{T}^x + i S_{T}^y}{2 \sqrt{2}} + \frac{S_{LT}^x + i S_{LT}^y}{2 \sqrt{2}} 
   & \frac{1}{3} - \frac{ 2 S_{LL} }{3}
   & \frac{S_{T}^x - i S_{T}^y}{2 \sqrt{2}} -\frac{S_{LT}^x - i S_{LT}^y}{2 \sqrt{2}} 
    \\[+0.20cm]
     \frac{S_{TT}^{xx} + i S_{TT}^{xy}}{2}
   & \frac{S_{T}^x + i S_{T}^y}{2 \sqrt{2}} -\frac{S_{LT}^x + i S_{LT}^y}{2 \sqrt{2}} 
   & \frac{1}{3}  - \frac{S_L}{2} + \frac{S_{LL}}{3}
    \end{array}
\right) .
\label{eqn:spin-1-spin-density}
\end{align}
The covariant forms of $S^\mu$ and $T^{\mu\nu}$ are generally expressed
by the longitudinal and transverse polarizations as 
\cite{bacchetta-2000-PRD,Boer-2016,vonDaal-2016}
\begin{align}
S^\mu & = S_L \frac{p^+}{M} \bar n^\mu - S_L \frac{M}{2  p^+} n^\mu + S_T^\mu ,
\nonumber \\
T^{\mu\nu} & = \frac{1}{2} \left [ \frac{4}{3} S_{LL} \frac{(p^+)^2}{M^2} 
               \bar n^\mu \bar n^\nu 
               + \frac{p^+}{M} \bar n^{\{ \mu} S_{LT}^{\nu \}}
               - \frac{2}{3} S_{LL} ( \bar n^{\{ \mu} n^{\nu \}} -g_T^{\mu\nu} )
+ S_{TT}^{\mu\nu} - \frac{M}{2 p^+} n^{\{ \mu} S_{LT}^{\nu \}}
+ \frac{1}{3} S_{LL} \frac{M^2}{(p^+)^2}n^\mu n^\nu \right ],
\label{eqn:spin-1-tensor-1}
\end{align}
where 
$a^{\{ \mu} b^{\nu \}}$ indicates the symmetrization of 
the superscript indices: $a^{\{ \mu} b^{\nu \}} \equiv a^\mu b^\nu + a^\nu b^\mu$,
and $g_T^{\alpha\beta}$ is defined by 
$g_T^{\alpha\beta} = g^{\alpha\beta} - \bar n^{\{ \alpha} n^{\beta \}}$
($g_T^{11}=g_T^{22}=-1$, $\text{others}=0$).
The light-like vectors $\bar n$ and $n$ are defined by
\begin{align}
\bar n^\mu  = \frac{1}{\sqrt{2}}  (\, 1,\, 0,\, 0,\, +1\, )
             = [\, 1,\, 0,\, \vec 0_T \, ]_{\text{LC}} , \ \ \ 
     n^\mu  = \frac{1}{\sqrt{2}}  (\, 1,\, 0,\, 0,\, -1\, )
             = [\, 0,\, 1,\, \vec 0_T \, ]_{\text{LC}} ,
\label{eqn:n-nbar}
\end{align}
where the lightcone (LC) notation 
$a^\mu = [\, a^+,\, a^-,\, \vec a_T \, ]_{\text{LC}}$
is used with $a^\pm = (a^0 \pm a^3)/\sqrt{2}$.
Replacing $S_{LL}$ in Ref.\,\cite{bacchetta-2002-PhD} by $-(2/3) S_{LL}$
so as to agree with the convention of 
Refs.\,\cite{bacchetta-2000-PRD,Boer-2016,vonDaal-2016}, we have
\cite{bacchetta-2002-PhD}
\begin{align}
S^\mu  =
\left[ \, \ S_L \frac{p^+}{M}, \ -S_L \frac{M}{2p^+},  \ S_T^x, \ S_T^y \,
\right]_{\text{LC}} , \ \ \ 
T^{\mu\nu}   = \frac{1}{2} 
\left[
    \begin{array}{cccc}
      \frac{4 (P^+)^2}{3 M^2} S_{LL}  & -\frac{2}{3} S_{LL}
   &  \frac{p^+}{M} S_{LT}^x          & \frac{p^+}{M} S_{LT}^y            \\[+0.20cm]
      -\frac{2}{3} S_{LL}             & \frac{M^2}{3 (p^+)^2} S_{LL}
   &  -\frac{M}{2 P^+} S_{LT}^x       & -\frac{M}{2 p^+} S_{LT}^y          \\[+0.20cm]
     \frac{p^+}{M} S_{LT}^x           & -\frac{M}{2 p^+} S_{LT}^x 
   &   S_{TT}^{xx}-\frac{2}{3} S_{LL} & S_{TT}^{xy}                        \\[+0.20cm]
      \frac{p^+}{M} S_{LT}^y          & -\frac{M}{2 p^+} S_{LT}^y  
   &   S_{TT}^{xy}                    & -S_{TT}^{xx}-\frac{2}{3} S_{LL}    
    \end{array}
\right]_{\text{LC}} .
\label{eqn:spin-1-Tmunu}
\end{align}
The lightcone matrix notation means that
the first component of column 
or line is $a^+$, the second is $a^-$, 
the third is the transverse coordinate $x$, 
and the fourth is the transverse $y$.
\end{widetext}

\begin{figure}[t]
\vspace{-0.00cm}
\begin{center}
   \includegraphics[width=4.0cm]{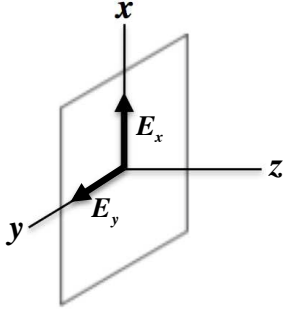}
\end{center}
\vspace{-0.6cm}
\caption{Linear polarizations $E_x$ and $E_y$ of spin-1 deuteron.}
\label{fig:linear-pol}
\vspace{-0.0cm}
\end{figure}

For investigating the gluon transversity distribution in the deuteron, 
the linear polarization of the spin-1 deuteron should be considered. 
The deuteron is linearly-polarized if its polarization
is $E_x$, $E_y$, or in between as illustrated 
in Fig.\,\ref{fig:linear-pol}.
In the expression of the general density matrix, 
the linear polarization asymmetry $E_x- E_y$ corresponds
to the spin asymmetry ${S_{TT}^{\alpha\beta}}$ 
in Refs.\,\cite{bacchetta-2000-PRD,bacchetta-2002-PhD,Boer-2016,vonDaal-2016}.
Therefore, the linear polarization asymmetry of the cross section
is calculated by taking $[d\sigma (E_x) - d\sigma (E_y)]/2$ with
the polarization $S_{TT}^{xx}$
and by taking other polarization as zero.
Later, the factor of 2 is multiplied in expressing Eq.\,(\ref{eqn:contribute-pdfs})
because the spin asymmetry $d\sigma (E_x) - d\sigma (E_y)$ is used
for expressing the cross section, instead of $[d\sigma (E_x) - d\sigma (E_y)]/2$.
As given in Eq.\,(\ref{eqn:spin-1-tensor-1}), the spin asymmetry $S_{TT}^{\mu\nu}$ 
is obtained from the tensor $T^{\mu\nu}$, by terminating other spin asymmetries
$S_{LL}=S_{LT}^x=S_{LT}^y=S_{TT}^{xy}=0$. Then, using the matrix form of
Eq.\,(\ref{eqn:spin-1-Tmunu}), we obtain
\begin{align}
S_{TT}^{\alpha\beta} 
= 
\left(
    \begin{array}{cccc}
      0    &    0   &  0     &  0         \\
      0    &   -1   &  0     &  0         \\
      0    &    0   &  1     &  0         \\
      0    &    0   &  0     &  0         \\
    \end{array}
\right) ,
\label{eqn:S_TT^xx}
\end{align}
where $S_{TT}^{xx} =-1$ is assigned for the linear polarization asymmetry
$E_x-E_y$, instead of $E_y-E_x$
in Refs.\,\cite{bacchetta-2000-PRD,bacchetta-2002-PhD,Boer-2016,vonDaal-2016}.

Next, we explain linear gluon polarizations which 
are related to the gluon transversity of the deuteron.
As shown in Eq.\,(\ref{eqn:delta-deltaT-qx-amplitudes}),
the transversity distribution is defined by the process
with the gluon polarizations $\lambda_i=+1$ and $\lambda_f=-1$ or vice versa. 
Therefore, the gluon polarization tensor $\varepsilon_{TT}^{ \alpha \beta}$
is given by the polarization vectors defined in Eq.\,(\ref{eqn:dct}) as
\begin{align}
\varepsilon_{TT}^{ \alpha \beta} & 
\equiv 
     \varepsilon_x^\alpha \varepsilon^{*\beta}_x
      -\varepsilon_y^\alpha \varepsilon^{*\beta}_y 
= - ( \varepsilon_+^\alpha  \varepsilon^{*\beta}_- 
       + \varepsilon_-^\alpha \varepsilon^{*\beta}_+ ) .
\label{eqn:gluon-polarization-tensor-1}
\end{align}
This relation indicates the helicity flip of 2 in this process. 
For investigating the gluon transversity, the parent hadron, 
namely the deuteron in this work, should have spin larger than 
or equal to one.
We notice that this gluon linear-polarization tensor 
$\varepsilon_{TT}^{ \alpha \beta}$ is the same as the 
linear-polarization asymmetry $S_{TT}^{\alpha\beta}$ in 
Eq.\,(\ref{eqn:S_TT^xx}) except for the sign:
\begin{align}
\varepsilon_{TT}^{ \alpha \beta} 
= - S_{TT}^{\alpha\beta} .
\label{eqn:linear-polarizations-g-d}
\end{align}
The tensor $S_{TT}^{\alpha\beta}$ (or $\varepsilon_{TT}^{ \alpha \beta}$)
appears in calculating the Drell-Yan production cross section
for finding the gluon transversity.

\subsection{Kinematical variables}
\label{kinematics}

\begin{figure}[b]
\vspace{-0.00cm}
\begin{center}
   \includegraphics[width=8.5cm]{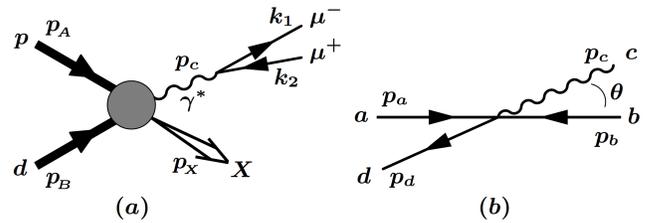}
\end{center}
\vspace{-0.6cm}
\caption{
$(a)$ Proton-deuteron Drell-Yan process
$p+d \to \mu^+\mu^- +X$.
$(b)$ Parton reaction $a+b \to c+d$ in the center-of-momentum frame.}
\label{fig:pd-Drell-Yan}
\vspace{-0.0cm}
\end{figure}

We express the cross section for 
the polarized proton-deuteron Drell-Yan process
($p + d \to \mu^+ \mu^- + X$),
which is illustrated in Fig.\,\ref{fig:pd-Drell-Yan}\,$(a)$,
in terms of the PDFs of the proton
and the deuteron 
\cite{drell-yan-books}
including the gluon transversity distribution.
The cross section for $A \, (p) + B\, (d) \to \mu^+ \mu^- + X$ is 
described by the partonic subprocess $\sigma_{ab \to cd}$
in Fig.\,\ref{fig:pd-Drell-Yan}\,$(b)$.
The indices $A$ and $B$ indicate the proton ($p$) 
and the deuteron ($d$), and $c$ is used for the virtual photon
($\gamma^*$).

Here, kinematical variables are explained for describing the process.
First, the Mandelstam variables for the reaction $p+d \to \gamma^*+X$
are given by neglecting proton and deuteron masses as
\begin{align}
s & = (p_A + p_B)^2 = (p_c + p_X)^2 = 2 p_A \cdot p_B ,
\nonumber \\[-0.00cm]
t & = (p_A - p_c)^2 = (p_B - p_X)^2 = Q^2 - 2 p_A \cdot p_c ,
\nonumber \\[-0.00cm]
u & = (p_A - p_X)^2 = (p_B - p_c)^2 = Q^2 - 2 p_B \cdot p_c ,
\label{eqn:mandel-1}
\end{align}
where $Q^2$ is defined by $Q^2  = p_c^2$.
It should be noted that the center-of-mass energy squared $s$ used
in the Fermilab-E906 experimental proposal is different from our definition.
Considering interactions with individual nucleons within the deuteron
\cite{Fermilab-dy,wen-chen-discuss}, they defined $s$ as
\begin{align}
s_{_{\text{Fermilab-E906}}} = \left ( p_p + \frac{p_d}{2} \right )^2 = \frac{s}{2} ,
\label{eqn:e906}
\end{align}
for the proton-deuteron Drell-Yan process as written in their proposal. 
One needs to be careful about the definition
difference in estimating the cross section numerically.
This difference is also related to how to define scaling variables 
for partons in the deuteron, and the details of handling this difference
are discussed in Sec.\,\ref{results} for showing numerical results.

In the proton-deuteron center-of-momentum (c.m.) frame, 
we denote the momenta as
\begin{align}
\! \! \! \! 
p_A  & =  \frac{\sqrt{s}}{2} \left ( \, 1,\, 0,\,  0,\,  1 \, \right ), \ \ 
p_B   = \frac{\sqrt{s}}{2} \left ( \, 1,\, 0,\,  0,\, -1 \, \right ), 
\nonumber \\[-0.00cm]
q & \equiv p_c   = 
      \left ( \, E,\, q_T \cos\phi,\,  q_T \sin\phi,\,  q_L \right )  ,
\nonumber \\[-0.00cm]
\ p_d &  =
|\vec q \,| \left ( \, 1,\,  -\sin\theta \cos\phi,\,
                     -\sin\theta \sin\phi,\,  -\cos\theta\, \right ) ,
\label{eqn:p-ABc}
\end{align}
where 
the polar and azimuthal angles of the virtual-photon momentum are given by $\theta$ 
and $\phi$, respectively, $|\vec q\,|$ is the photon momentum, 
and $q_L = |\vec q\, | \cos\theta$ and $q_T = |\vec q\, | \sin\theta$ 
are longitudinal and transverse momenta of photon.
The $z$ direction is taken as the momentum direction for the proton
($z \parallel \vec p_A$).
The dimuon momentum is equal to the virtual photon momentum:
\begin{align}
q = k_1 + k_2 ,
\label{eqn:dimuon-momentum}
\end{align}
where $k_1$ and $k_2$ are $\mu^-$ and $\mu^+$ momenta,
so that the scale $Q^2$ is the dimuon-mass squared:
\begin{align}
M_{\mu\mu}^2 = (k_1 + k_2)^2 =Q^2 .
\label{eqn:dimuon-mass}
\end{align}
The dimensionless variable $\tau$ and the dimuon rapidity $y$
are defined by
\begin{align}
\tau = \frac{Q^2}{s} ,
\label{eqn:tau}
\end{align}
and
\begin{align}
y = \frac{1}{2} \ln \frac{E+q_L}{E-q_L} 
  = - \ln \left [ \tan (\theta/2)\right ] ,
\label{eqn:rapidity}
\end{align}
where $y$ is given by the energy $E$ and momentum $q_L$ 
in the c.m. frame.
The photon momentum is also expressed by using the rapidity,
the transverse momentum, and the transverse mass as
\begin{align}
q  = \left ( \, M_T \cosh y,\,  q_T \cos\phi,\,  
           q_T \sin\phi,\, M_T \sinh y \, \right ) 
\label{eqn:photon-q-rapidity}
\end{align}
where the transverse mass is given by
\begin{align}
M_T = \sqrt{Q^2 + \vec q_T^{\,\, 2}} .
\label{eqn:transverse-mass}
\end{align}
In addition, the variables $x_1$ and $x_2$ are defined by
\cite{drell-yan-books}
\begin{align}
x_1 & = - \frac{u-Q^2}{s} 
= \frac{M_T}{\sqrt{s}} \, e^{\,y} ,
\nonumber \\
x_2 & = - \frac{t-Q^2}{s} 
=  \frac{M_T}{\sqrt{s}} \, e^{-y} ,
\label{eqn:x1-x2}
\end{align}
for describing the Drell-Yan cross section. 

Next, kinematical variables are shown for parton reactions
of Fig.\,\ref{fig:pd-Drell-Yan}$(b)$.
The initial parton momenta are given by the momentum fractions
of partons, $x_a$ and $x_b$, with respect to their parent-hadron
momenta as:
\begin{align}
p_a & = x_a P_A = x_a \frac{\sqrt{s}}{2} \left ( \, 1,\, 0,\,  0,\,  1 \, \right ), 
\nonumber \\[-0.00cm]
p_b & = x_b P_B = x_b \frac{\sqrt{s}}{2} \left ( \, 1,\, 0,\,  0,\, -1 \, \right ). 
\label{eqn:p-ab}
\end{align}
Then, the Mandelstam variables in the partonic level 
for the reaction $a+b \to c+d$ are expressed as
\begin{align}
\hat s & = (p_a + p_b )^2 = 2 p_a \cdot p_b = x_a x_b s ,
\nonumber \\[-0.00cm]
\hat t & = (p_a - p_c)^2 
         = Q^2 - 2 p_a \cdot q = Q^2 + x_a (t-Q^2) ,
\nonumber \\[-0.00cm]
\hat u & = (p_b - p_c)^2 
         = Q^2 - 2 p_b \cdot q = Q^2 + x_b (u-Q^2) .
\label{eqn:mandel-parton}
\end{align}

Since the gluon transversity of the deuteron is studied 
in this work, we consider that the deuteron is linearly polarized
with the polarization vectors $E_x$ and $E_y$ in Eq.\,(\ref{eqn:dct}).
In probing the gluon transversity of the deuteron by the reaction
$p + d \to \mu^+ \mu^- + X$, this deuteron polarization needs to be considered
with the unpolarized proton beam.
In particular, the following combination of the polarized differential 
cross sections should be studied:
\begin{align}
& d\sigma(E_x) -d\sigma(E_y)   .
\label{eqn:polarized-cross-1}
\end{align}
Then, the linear polarization tensor of Eqs.\,(\ref{eqn:S_TT^xx}),
(\ref{eqn:gluon-polarization-tensor-1}), and
(\ref{eqn:linear-polarizations-g-d}) 
appears in calculating the cross section asymmetry.
Later, there appears the following contraction
\begin{align}
q_\mu \, S_{TT}^{\mu\nu} \, q_\nu = - q_T^2 \cos (2\phi) ,
\label{eqn:eps-contraction-1}
\end{align}
in calculating the cross section.
Equation (\ref{eqn:eps-contraction-1}) is obtained by using
the momentum assignments of Eq.\,(\ref{eqn:p-ABc}).

\subsection{Parton correlation and distribution functions\\ 
            of proton and deuteron}
\label{pd-pdfs}

\begin{figure}[b]
 \vspace{-0.00cm}
\begin{center}
   \includegraphics[width=6.0cm]{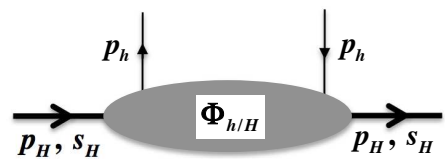}
\end{center}
\vspace{-0.5cm}
\caption{Quark correlation function $\Phi_{h/H}$.}
\label{fig:parton-correlation}
\vspace{-0.30cm}
\end{figure}

The cross section for $p + d \to \mu^+ \mu^- + X$ is expressed
in terms of parton correlation functions and subsequently
by collinear PDFs, so that they are introduced in this section.
Later, the cross section is expressed by the hadron tensor $W_{\mu\nu}$
multiplied by the photon tensor $L^{\mu\nu}$ in Sec.\,\ref{cross}.
The hadron tensor contains parton correlation functions in
the proton and deuteron.
There are correlation functions for quarks, antiquarks, and gluon
($h=q,\bar q, g$) in the hadron $H$:
\begin{align}
\! \! 
\Phi_{h/H} = \Phi_{u/H}  , \, \Phi_{d/H} , \, \cdots , \, 
\Phi_{\bar u/H} , \, \Phi_{\bar d/H} , \, \cdots , 
  \, \Phi_{g/H} .
\label{eqn:correlation-sum}
\end{align}
The quark correlation functions are illustrated in 
Fig.\,\ref{fig:parton-correlation}
and they are defined by the matrix elements for quarks as
\begin{align}
& 
\Phi_{q/H}^{\, ij} (p_h, p_H, s_H)
\nonumber \\[-0.00cm]
& \! \! \! \! 
=
\int  \! \frac{d^4 \xi_h}{(2\pi)^4} \, e^{i p_h \cdot \xi_h}
\langle \, p_H s_H \left | \, 
\bar\psi _j (0) \, \psi _i (\xi_h)  \, \right | p_H s_H \, \rangle  ,
\label{eqn:correlation-q}
\end{align}
where $\psi$ is the quark field, and $\xi_h$ is 
a four-dimensional space-time coordinate.
To be precise, the gauge link should exist
between $\bar\psi _j (0)$ and  $\psi _i (\xi_h) $
to satisfy the color gauge invariance; however,
they are not explicitly written in this article.
The correlation function indicates
the amplitude to extract a parton from a hadron 
and then to insert it into the hadron
at a different spacetime point.
The correlation function for the antiquark is given
by changing 
$ \bar\psi _j (0) \psi _i (\xi_h) $ for
$ \psi _i (0) \bar\psi _j (\xi_h) $:
\begin{align}
& 
\Phi_{\bar q/H}^{\, ij} (p_h, p_H, s_H)
\nonumber \\[-0.00cm]
& \! \! \! \! 
= \int  \! \frac{d^4 \xi_h}{(2\pi)^4}
\, e^{i p_h \cdot \xi_h}
\langle \, p_H s_H \left | \, \psi _i (0) \, \bar\psi _j (\xi_h) 
\, \right | p_H s_H \, \rangle  .
\label{eqn:correlation-qbar}
\end{align}
The quark and antiquark correlation functions in 
Refs.\,\cite{bacchetta-2000-PRD},
\cite{Boer-2016}, and \cite{vonDaal-2016} are the same as 
the one in Eqs.\,(\ref{eqn:correlation-q})
and (\ref{eqn:correlation-qbar}).
However, they are defined without the $1/(2\pi)^4$ factor
in Ref.\,\cite{br-book}.

The gluon correlation function is defined in the similar way by
\begin{align}
\Phi_{g/H}^{\, \alpha\beta} (p_h, p_H, s_H) &
= N_{g/H} \int \! \frac{d^4 \xi}{(2\pi)^4}
\, e^{i p_h \cdot \xi_h}
\nonumber \\[-0.10cm]
& \ 
\times
\langle \, p_H s_H \! \left | \, 
A^{\alpha} (0) A^{\beta} (\xi)  \, \right |  p_H s_H \, \rangle  ,
\label{eqn:correlation-g-1}
\end{align}
where $A^\alpha$ is given by $A^\alpha = A^\alpha_a \, t^a$
as explained below Eq.\.(\ref{eqn:delta-deltaT-gx}),
and $N_{h/H}$ is the normalization constant.
The gluon correlation function is often expressed 
by the gluon-field strength tensor 
$F^{\mu\nu}_a = \partial^\mu A^\nu_a 
    - \partial^\nu A^\mu_a + g f_{abc} A^\mu_b A^\mu_c$
\cite{ji-1992,br-book,Boer-2016,vonDaal-2016}.
Let us consider the hadron $H$ with momentum 
in the positive-$z$ direction.
In the lightcone gauge $A^+ =0$, the gluon filed has
three components consist of $\vec A_\perp$ and $A^-$,
and it satisfies $ \partial^+ A^\mu =F^{+\mu}$.
Therefore, Eq.\,(\ref{eqn:correlation-g-1}) becomes
\begin{align}
& \! \! \! \! 
\Phi_{g/H}^{\, \alpha\beta} (p_h,\, p_H,\, s_H)
= \frac{1}{N_{g/H}} 
 \int \! \frac{d^4 \xi} {(2\pi)^4}
 \, e^{i p_h \cdot \xi_h}
\nonumber \\[-0.10cm]
& \ \ \ \ \ \ \ \ \ \ \ \ \ \ \ \ 
\times  \langle \, p_H s_H \left | \, 
F^{\, + \alpha} (0) F^{\, + \beta} (\xi_h) 
 \, \right | p_H s_H \, \rangle  ,
\label{eqn:correlation-g-2}
\end{align}
where $F^{\, \mu \nu}$ is defined with the color factor as
$F^{\, \mu \nu} = F^{\, \mu \nu}_a \, t^a$.
One may note that the gluon correlation function 
in Eq.\,(\ref{eqn:correlation-g-2})
is slightly different from the ones in Refs.\,\cite{Boer-2016,vonDaal-2016}
by the factor of $1/N_{g/H}$:
$\Phi_{g/H}^{\, \alpha\beta} (p_h, p_H, s_H)_{\text{our}}
= (1 / N_{g/H})
\Phi_{g/H}^{\, \alpha\beta} 
(p_h, p_H, s_H)_{\text{\cite{Boer-2016,vonDaal-2016}}}$.
The overall normalization constant $N_{h/H}$ is different
depending on the hadron momentum direction, namely 
in the positive- or negative-$z$ direction:
\begin{align}
\! \! \!
N_{g/H} = \left\{
   \begin{aligned}[l]
       p_h^+ \  & \text{\ \ \  for $H=A$} \\
       p_h^- \  & \text{\ \ \  for $H=B$} 
   \end{aligned} 
   \right. .
\label{eqn:correlation-norm}
\end{align}
Here, $p_h^\pm$ are the lightcone momenta.
The details of this gluon normalization factor are explained
in Ref.\,\cite{collins-spin-book}.

Next, we define TMDs and the collinear PDFs from the correlation functions.
The correlation functions are integrated 
over $p_h$ ($p_a$ or $p_b$) to obtain 
the collinear correlation functions:
\begin{align}
& \Phi_{h/H} (x_h) = \int \! d^2 p_{hT} \, \Phi_{h/H} (x_h, p_{hT})
\nonumber \\
& \ \ \ 
= \int \! d^4 p_{h} \, 
\Phi_{h/H} (p_h, p_H, s_H) \, 
\delta (p_h^\pm -x_h p_H^\pm) ,
\label{eqn:correlation-parton-2}
\end{align}
where $\vec p_{hT}$ is the transverse momentum of the parton $h$,
and $\Phi_{h/H} (x_h, \vec p_{hT})$ are
transverse-momentum-dependent correlation functions, which
are related to the TMDs.
The $+/-$ indicates the $\delta$ function
$\delta (p_h^{\scriptscriptstyle{+}}-x_h p_A^{\scriptscriptstyle{+}})$
in the proton $A$ or
$\delta (p_h^{\scriptscriptstyle{-}}-x_h p_B^{\scriptscriptstyle{-}})$
in the deuteron $B$.
The notation $\pm$ indicates $+$ ($-$) for $h=a$ and $H=A$ ($h=b$ and $H=B$).

There are some differences from other publication 
in kinematical factors
\cite{bacchetta-2000-PRD,br-book,Boer-2016,vonDaal-2016}.
The relations are  
\begin{align}
& \Phi_{q/A} (p_q, p_A, s_A)
= \Phi_{q/A} (p_q, p_A, s_A)
_{\text{\cite{bacchetta-2000-PRD,br-book,vonDaal-2016}}} , 
\nonumber \\
& \Phi_{q/A} (x_a)
  = \Phi_{q/A}  (x_a)_{\text{\cite{vonDaal-2016}}} 
= \frac{1}{p_A^+} \Phi_{q/A}  (x_a)
 _{\text{\cite{bacchetta-2000-PRD,br-book}}} ,
\label{eqn:correlation-fun-differences-q}
\end{align}
by taking $H=A$ for the quark correlation functions and the PDFs. 
The TMD relations are the same as the ones for the above PDFs.
In the same way, there are differences for the gluon correlation
function and its collinear distribution function as
\begin{align}
& \Phi_{g/A}^{\, \alpha\beta} (p_g, p_A, s_A)
 = \frac{1}{p_{g/A}^+} \,
\Phi_{g/A}^{\, \alpha\beta} (p_g, p_A, s_A)
_{\text{\cite{bacchetta-2000-PRD,Mulders-2001-PRD,Boer-2016,vonDaal-2016}}} , 
\nonumber \\
& \Phi_{g/A}^{\, \alpha\beta} (x_a)
  =  \frac{1}{x_a \, p_A^+} \,
\Phi_{g/A}^{\, \alpha\beta}  (x_a)_{\text{\cite{vonDaal-2016}}} 
= \frac{1}{x_a} \, \Phi_{g/A}^{\, \alpha\beta}  (x_a)
_{\text{\cite{bacchetta-2000-PRD,Mulders-2001-PRD,Boer-2016}}} , 
\label{eqn:correlation-fun-differences-g}
\end{align}

\vspace{-0.25cm}
\noindent
where $p_{g/A}^+ = x_a p_A^+$ is used.
Namely, the collinear correlation functions are defined as 
dimensionless quantities in this article.

In this work, the collinear functions are considered for finding
the gluon transversity, so that the parton momenta are integrated 
except for the $+$ or $-$ lightcone component
as given in Eq.\,(\ref{eqn:correlation-parton-2}). 
Usually, all the possible distribution functions
are listed in the TMD correlation-function form, 
so that the TMDs $f_{h/H} (x, \vec p_{hT})$ are integrated over $\vec p_{hT}$
to become the PDFs $f_{h/H} (x)$ for our work:
\begin{align}
f_{h/H} (x) & = \int \! d^2 p_{hT} \, f_{h/H} (x, \vec p_{hT}) .
\label{eqn:correlation-integrated}
\end{align}
The quark and gluon correlation functions are expressed
by the TMDs for the proton and deuteron 
\cite{br-book,bacchetta-2000-PRD,Boer-2016,vonDaal-2016},
and they are integrated over $\vec p_{hT}$.
Here, we are interested in probing the gluon transversity 
distribution of the deuteron. For finding it in the collinear formalism,
we first consider the leading-twist part
\cite{br-book}:
\begin{align}
\Phi_{q/A} (x_a) 
= \frac{1}{2} \bigg [ & \, 
 \slashed{\bar n} \, f_{1,q/A} (x_a) 
\! + \! \gamma_5 \, \slashed{\bar n} \, S_{A,L} \, g_{1,q/A} (x_a)
\nonumber \\[-0.10cm]
& \ \ 
 +    \slashed{\bar n} \, \gamma_5 \, \slashed{s}_{A,T}  h_{1,q/A} (x_a) 
\bigg ] ,
\label{eqn:correlation-integrated-q-proton} 
\end{align}
as explained in Sec.\,\ref{cross}.
Here, $f_{1,q/A} (x_a)$, $g_{1,q/A} (x_a)$, and $h_{1,q/A} (x_a)$ 
($\equiv \Delta_T q(x_a)$ in this paper)
are twist-2 distribution functions which indicate
unpolarized, longitudinally-polarized, and transversity distributions.

For the spin-1 deuteron, one should note that there are additional 
structure functions in comparison with the spin-1/2 nucleon ones 
due to the spin-1 nature.
The twist-2 part of the quark correlation function for the spin-1 deuteron is
given as \cite{bacchetta-2000-PRD,vonDaal-2016}
\begin{align}
\Phi_{q/B} & (x_b) 
 = \frac{1}{2} \bigg [  \, 
\slashed{n}  \, f_{1,q/B} (x_b) 
+ \gamma_5 \, \slashed{n} \, S_{B,L} \, g_{1,q/B} (x_b)
\nonumber \\
& 
+ \slashed{n} \, \gamma_5 \, \slashed{s}_{B,T} \, h_{1,q/B} (x_b)
  + \slashed{n} \, S_{B,LL} \, f_{1 LL, q/B} (x_b)
\nonumber \\
& 
+ \sigma_{\mu\nu} \, n ^\nu \, S_{B,LT}^\mu \, h_{1LT,q/B} (x_b) \,
\bigg ] ,
\label{eqn:correlation-integrated-q-deuteron} 
\end{align}
where $f_{1,q/B} (x_b)$ is the unpolarized distribution function,
$g_{1,q/B} (x_b)$ is the longitudinally-polarized one,
$h_{1,q/B} (x_b)$ ($= \Delta_T q(x_b)$)
is the transversity,
and $f_{1 LL, q/B} (x_b)$ and $h_{1LT,q/B} (x_b)$ 
are tensor-polarized ones of the deuteron.
The last two terms with $S_{B,LL}$ and $S_{B,LT}$ exist due to
the spin-1 nature of the deuteron, and they do not exist
for the spin-1/2 proton.
The correlation functions and the PDFs for antiquarks are
obtained by the replacement $q \to \bar q$ in these expressions.

For the gluon correlation function in the deuteron, 
the twist-2 part is similarly given as \cite{Boer-2016,vonDaal-2016}
\begin{align}
& \! \! \! \! 
\Phi_{g/B}^{\, \alpha\beta} (x_b) 
\equiv \int d^2 p_{bT} \, \Phi_{g/B}^{\, \alpha\beta} (x, \vec p_{bT}) \,
\nonumber \\
& \! \! \! \! \! 
= \frac{1}{2} \bigg [ - g_T^{\, \alpha\beta} \, f_{1,g/B} (x_b) 
+ i \, \epsilon_T^{\, \alpha\beta} \, S_{B,L} \, g_{1,g/B} (x_b)
\nonumber \\
& \! \! 
- g_T^{\, \alpha\beta} \, S_{B,LL} \, f_{1LL,g/B} (x_b) 
+ S_{B,TT}^{\, \alpha\beta} \, h_{1TT,g/B} (x_b)
\, \bigg ] ,
\label{eqn:correlation-integrated-g} 
\end{align}
where $f_{1,g/B}$ is the unpolarized gluon distribution function,
$g_{1,g/B}$ is the longitudinally-polarized one,
$f_{1 LL, g/B}$ and 
$h_{1TT,g/B}$ \cite{Boer-2016,vonDaal-2016}
($\equiv - \Delta_T g_B$ in this article)
are tensor- and linearly-polarized ones. 
We have a negative sign for $\Delta_T g_B$
because the linear polarization for the gluon transversity
is defined by $\varepsilon_x -\varepsilon_y$ 
in Eq.\,(\ref{eqn:gluon-polarization-tensor-1})
instead of $\varepsilon_y -\varepsilon_x$ 
in Refs.\,\cite{Boer-2016,vonDaal-2016}.
There are different definitions on the gluon transversity 
and its notation:
\begin{alignat}{2}
\Delta_2 G (x) & = g_{\hat x/ \hat x} (x) - g_{\hat y/ \hat x} (x)  & \ \ \ &
\text{\cite{artru-mekhfi,Mulders-2001-PRD}},
\nonumber \\
a (x) & = g_{\hat x/ \hat x} (x) - g_{\hat y/ \hat x} (x)           & \ \ \ &
\text{\cite{jlab-gluon-trans,transversity-model}},
\nonumber \\
\Delta_L g (x) & = g_{\hat x/ \hat x} (x) - g_{\hat y/ \hat x} (x)  & \ \ \ &
\text{\cite{transversity-q2-gluon}},
\nonumber \\
\delta G (x) & = - g_{\hat x/ \hat x} (x) + g_{\hat y/ \hat x} (x)  & \ \ \ &
\text{\cite{ss-1990,transversity-lattice}},
\nonumber \\
h_{1TT,g} (x) & = - g_{\hat x/ \hat x} (x) + g_{\hat y/ \hat x} (x) & \ \ \ &
\text{\cite{bacchetta-2000-PRD,Boer-2016,meissner-2007}},
\nonumber \\
\Delta_T g (x) & = g_{\hat x/ \hat x} (x) - g_{\hat y/ \hat x} (x) & \ \ \ &
\text{\cite{sterman-qcd}},\ \text{this work},
\label{eqn:gluon-transversity-definition} 
\end{alignat}
where $g_{\,\hat h/\hat H}$ indicates the gluon distribution with
the gluon polarization $\hat h$ and the hadron (deuteron) 
polarization $\hat H$.
In Eq.\,(\ref{eqn:correlation-integrated-g}), the notation 
$\epsilon_T^{\alpha\beta}$ is defined by
$\epsilon_T^{\alpha\beta} \equiv \epsilon^{\alpha\beta - +}$
($\epsilon_T^{12}=-\epsilon_T^{21}=1$, $\text{others}=0$).
It is different from the linear-polarization tensor
$\varepsilon_{TT}^{\alpha\beta}$.
The gluon correlation function $\Phi_{g/B}^{\, \alpha\beta} (x)$ is
expressed by only the transverse indices $\alpha$ and $\beta$ ($=1,\,2$),
because other components do not contribute to the cross section
in the leading-twist level.
The gluon correlation function $\Phi_{g/A}^{\alpha\beta} (x)$ of the proton 
is the same as the one for the deuteron 
in Eq.\,(\ref{eqn:correlation-integrated-g})
by terminating the tensor terms ($S_{LL}=S_{TT}^{\, \alpha\beta}=0$).
However, its expression is not explicitly written
because it does not contribute to the Drell-Yan cross section
in our current linear-polarization asymmetry as the leading one.

\subsection{Cross section for Drell-Yan process
$\mathbold{p + d \to \mu^+\mu^- + X}$ in parton model}
\label{cross}

The cross section for the proton-deuteron  Drell-Yan process 
$p + d \to \mu^+\mu^- + X$ 
of Fig.\,\ref{fig:pd-Drell-Yan}\,$(a)$ is calculated
by the partonic-subprocess cross section $\hat\sigma_{ab \to cd}$ 
in Fig.\,\ref{fig:pd-Drell-Yan}\,$(b)$.
The partonic cross section is convoluted with the PDFs of hadrons,
as explained in Ref.\,\cite{drell-yan-books}
for the unpolarized cross section.
Especially, since the correlation-function formalism is used
in deriving the polarized cross section, we first show 
the Drell-Yan cross section in terms of the parton correlation functions. 

We investigate the dimuon production with finite $q_T$.
First, the $q (\text{in }p) \bar q (\text{in }d)$ subprocess contribution 
is considered as an example to confirm our formalism 
with the correlation functions.
The unpolarized $pd$ Drell-Yan cross section $d\sigma_{pd \to \gamma X}$ 
is expressed by the partonic cross section for
$q (p) \bar q (d) \to \gamma^* g \to \mu^+ \mu^- g$ 
by the convolution with the quark and antiquark distribution functions as
\begin{align}
\left.
d \sigma_{pd \to \mu^+ \mu^- X} 
\right |_{q\bar q \to \gamma^* g}
= \int_0^1 dx_a & \int_0^1 dx_b \, q (x_a) \, \bar q(x_b) \,
\nonumber \\
& \ \ \ \ 
\times
d \hat\sigma_{q \bar q \to \mu^+ \mu^- g} .
\label{eqn:cross-convolution}
\end{align}
The partonic cross section for 
$q \bar q \to \gamma^* g \to \mu^+ \mu^- g$
is given by
\begin{align}
d \hat\sigma_{q \bar q  \to \mu^+ \mu^- g} & = 
 \frac{1}{4 p_a \cdot p_b} 
 \sum_{\substack{\text{spin,}\\[-0.02cm] \text{color}}}^{\rule{0.5cm}{0.4pt}}
 \sum_{\text{flavor}} 
 \left | M_{q \bar q \to \gamma^* g \to \mu^+ \mu^- g} \right |^2 
\nonumber \\[-0.00cm]
& \ \ 
\times
 (2\pi)^4 \delta ^4 (p_a +p_b -k_1-k_2 -p_d)
\nonumber \\[-0.00cm]
& \ \ 
\times
\frac{d^3 k_1}{2E_1 (2\pi)^3}
\frac{d^3 k_2}{2E_2 (2\pi)^3}
\frac{d^3 p_d}{2E_d (2\pi)^3} ,
\label{eqn:cross-parton-1}
\end{align}
where $k_1$, $k_2$ and $p_d$ are momenta 
for $\mu^-$, $\mu^+$, and gluon in the final state.
The three-body phase space for the final state
is written by the two-body
phase spaces as \cite{pdg}
\begin{align}
d\Phi_3 (p_a & +p_b; k_1, k_2, p_d) 
\nonumber \\
&
= d\Phi_2 (q;k_1, k_2) \, d\Phi_2 (p_a+p_b;q, p_d) ,
\label{eqn:2-3-phase-spaces}
\end{align}
where the $n$-body phase space is defined by
\begin{align}
\! \!
d\Phi_n (P;p_1, \cdots, p_n)
& = \delta ^4 \! \left ( \! P - \! \sum_{i=1}^n p_i \right) 
\prod_{i=1}^n
   \frac{d^3 p_i}{2E_i (2\pi)^3} .
\label{eqn:n-phase-space}
\end{align}

The matrix element is described by the process
$q\bar q \to \gamma^* g$ and subsequent 
$\gamma^* \to \mu^+ \mu^-$ as
\begin{align}
& M_{q \bar q  \to \gamma^* g \to \mu^+ \mu^- g}
 = e \, M_{\gamma^* \to \mu^+ \mu^-}^\mu \frac{-1}{Q^2}
  \, e \, M_{q \bar q \to \gamma^* g, \,\mu} ,
\nonumber \\
& \ 
M_{\gamma^* \to \mu^+ \mu^-}^\mu \,
 = \bar u (k_1,\lambda_1) \gamma^\mu v(k_2,\lambda_2) ,
\nonumber \\
& \ 
M_{q \bar q \to \gamma^* g, \,\mu}
 = e_q \, \varepsilon^{* \alpha} (p_d, \lambda_d) \,
      \bar v (p_b,\lambda_b) \Gamma_{\mu\alpha} u(p_a,\lambda_a) ,
\label{eqn:qqbar-matrix}
\end{align}
where $\Gamma_{\mu\alpha}$ indicates the $q \bar q \gamma g$ 
interaction part.
The dimuon term is calculated and it becomes the lepton tensor
$L^{\mu\nu}$:
\begin{align}
\sum_{\lambda_1,\,\lambda_2} 
&
\left ( M_{\gamma^* \to \mu^+ \mu^-}^\mu  \right )^\dagger
 M_{\gamma^* \to \mu^+ \mu^-}^\nu 
\nonumber \\[-0.15cm]
& = 2 \, L^{\mu\nu} = 4  \, \left ( \, k_1^\mu k_2^\nu + k_1^\nu k_2^\mu 
                    - k_1 \cdot k_2 \, g^{\mu\nu} \, \right ) ,
\label{eqn:lepton-tensor}
\end{align}
where the overall factor of 2 is assigned so that 
the lepton tensor agrees with the conventional one 
used in describing the deep inelastic lepton-nucleon scattering.
Its integral over the phase space is given by
\begin{align}
\int \! d\Phi_2 (q;k_1, k_2) \, 2 \, L^{\mu\nu}
= \frac{1}{6 \pi} (q^\mu q^\nu -Q^2 g^{\mu\nu}) .
\label{eqn:phase-integral-Lmunu}
\end{align}
This relation eventually becomes $- Q^2 g^{\mu\nu}/(6 \pi)$
by considering the current conservation 
$q^\mu M_{q \bar q \to \gamma^* g, \,\mu} =0$.

From the relations in Eq.\,(\ref{eqn:mandel-parton}),
the four-momentum square $p_d^2$ becomes
\begin{align}
\! \! 
p_d^2 = (p_a +p_b -q)^2 
= s (x_a -x_1) \! \left[ x_b -\frac{x_a x_2 -\tau}{x_a-x_1} \right] ,
\label{eqn:pd2}
\end{align}
where the used kinematical variables ($x_1$, $x_2$, $\cdots$)
are defined in Sec.\,\ref{kinematics}.
Then, using the relations for the integrals
\begin{align}
& 
\frac{d^3 q}{2 E} = \frac{1}{4} \, dq_T^2 \, d\phi \, dy, \ \ 
\frac{d^3 p_d}{2 E_d} = d^4 p_d \, \delta (p_d^2), 
\nonumber \\
& 
\int_0^1 dx_b \, \delta (p_d^2) = \frac{1}{s(x_a-x_1)},
\label{eqn:integrals}
\end{align}
and taking the gluon-spin summation,
we finally obtain the cross-section expression as
\begin{align}
& \! \! \! \! \! \! \! \! 
\left.
\frac{ d \sigma_{pd \to \mu^+ \mu^- X} }{d\tau \, dq_T^2 \, d\phi \, dy}
\right |_{q\bar q \to \gamma^* g}
\! \! = \frac{\alpha^2}{12 (2\pi)^2 Q^2} 
  \sum_{\substack{\text{spin,}\\[-0.02cm] \text{color}}}^{\rule{0.5cm}{0.4pt}}
  \sum_q e_q^2 \,  g^{\mu\nu} g^{\alpha\beta}  
\nonumber \\
& \! \! \! \! \! \! \! 
\times \!
  \int_{\text{min}(x_a)}^1 \frac{1}{x_a-x_1}
q_A (x_a) \bar q_B (x_b) 
\text{Tr} \left [  \frac{\gamma^+}{2} \Gamma_{\nu\beta}
                  \frac{\gamma^-}{2} \hat \Gamma_{\mu\alpha}  \right ] .
\label{eqn:cross-parton-II}
\end{align}
Here, due to the $\delta$-function for $p_d^2$ in
Eqs.\,(\ref{eqn:pd2}) and (\ref{eqn:integrals}) 
and also the kinematical constraint of max($x_b$)=1, 
we have the relations
\begin{align}
x_b= \frac{x_a x_2 - \tau}{x_a - \tau}, \ \ 
\text{min}(x_a) = \frac{x_1-\tau}{1-x_2} ,
\label{eqn:min-xa}
\end{align}
and the notation $\hat O$ is defined by
\begin{align}
\hat O \equiv \gamma^0 \, O^\dagger \, \gamma^0 .
\label{eqn:gamma-bar}
\end{align}
The fine structure constant $\alpha$ is given by $\alpha=e^2/(4 \pi)$.

\begin{figure}[t]
\vspace{-0.00cm}
\begin{center}
   \includegraphics[width=6.0cm]{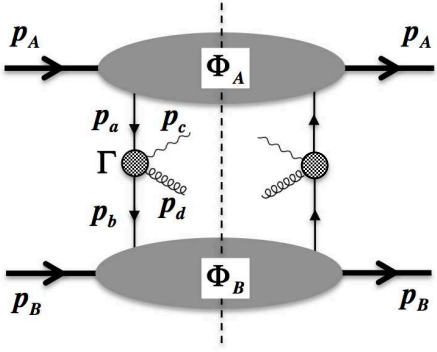}
\end{center}
\vspace{-0.6cm}
\caption{$q+\bar q \to \gamma^* +g$ process for cross section of 
$p+d \to \gamma^* +X$}
\label{fig:pd-parton-1}
\vspace{-0.20cm}
\end{figure}

\begin{widetext}
Next, we try to write the cross section 
in terms of parton correlation functions in Sec.\,\ref{pd-pdfs}
by considering only the subprocess 
$q(p) + \bar q (d) \to \gamma + g$
in Fig.\,\ref{fig:pd-parton-1} as 
\begin{align}
\left.
d\sigma_{p d \to \gamma X} \right |_{q\bar q \to \gamma^* g}
& = \frac{1}{4 \, p_A \cdot p_B} 
\int \! \frac{d^4 p_a}{(2\pi)^4} \int \! \frac{d^4 p_b}{(2\pi)^4}
\sum_{\substack{\text{spin,}\\[-0.02cm] \text{color}}}^{\rule{0.5cm}{0.4pt}}
\sum_{\text{flavor}} 
 \sum_{X_A,X_B} 
 (2 \pi)^4 \, \delta^4 (p_A-p_a-p_{AX}) 
 (2 \pi)^4 \, \delta^4 (p_B-p_b-p_{BX}) 
\nonumber \\
& \ \ \ 
\times
\left | \, \langle X_B \left | \bar \psi_{b,l} (0) \right | p_B s_B \rangle \,
            ( \Gamma_{q\bar q \to \gamma^* g, \, \mu} )_{lk} \,
           \langle X_A \left | \psi_{a,k} (0) \right | p_A s_A \rangle 
          \, M_{\gamma^* \to \mu^+ \mu^-}^\mu  \, \right |^2 
\nonumber \\
& \ \ \ 
\times
\left ( \frac{-e}{Q^2} \right )^2
  (2 \pi)^4 \, \delta^4 (p_a+p_b-k_1-k_2-p_d) 
 \frac{d^3 k_1}{2E_1 (2\pi)^3}\frac{d^3 k_2}{2E_2 (2\pi)^3}
 \frac{d^3 p_d}{2E_d (2\pi)^3} ,
\label{eqn:cross-parton-4}
\end{align}
where the spin summations are taken for muons, quark, antiquark, and gluon.
The parton-interaction part $\Gamma_{q\bar q \to \gamma^* g, \, \mu}$ 
is given by
$ \Gamma_{q\bar q \to \gamma^* g, \, \mu}
  = e_q \, \varepsilon^{* \alpha} (p_d, \lambda_d) \,
    \Gamma_{\mu\alpha}$
by extracting out the quark charge $e_q$ and the gluon-polarization vector 
$\varepsilon^{* \alpha} (p_d, \lambda_d)$ from $\Gamma_{q\bar q \to \gamma^* g, \, \mu}$. 
By changing the three-body phase space to the two-body ones and
repeating the same calculations from Eq.\,(\ref{eqn:lepton-tensor})
to Eq.\,(\ref{eqn:integrals}), the cross section is written by 
the lepton tensor $L^{\mu\nu}$ in Eqs.\,(\ref{eqn:lepton-tensor}) 
and (\ref{eqn:phase-integral-Lmunu})
multiplied by the hadron tensor $W_{\mu\nu}$:
\begin{align}
\frac{ d \sigma_{pd \to \mu^+ \mu^- X} }{d\tau \, dq_T^2 \, d\phi \, dy}
= 
\frac{\alpha^2}{2 (2\pi)^2 Q^4} 
\left ( \int \! d\Phi_2 (q;k_1, k_2) \, 2 \, L^{\mu\nu} \right)
\, W_{\mu\nu} .
\label{eqn:cross-parton-I}
\end{align}
This cross-section expression, in terms of lepton and hadron tensors,
can be used not only for the process $q\bar q \to\gamma^* g$ under consideration
but also for any partonic subprocesses in the Drell-Yan process.
The hadron tensor is given by
\begin{align}
& W_{\mu\nu} (q \bar q) = 
 \int \frac{d^4 p_a}{(2\pi)^4} \int \frac{d^4 p_b}{(2\pi)^4} \,
 \sum_{\substack{\text{spin,}\\[-0.02cm] \text{color}}}^{\rule{0.5cm}{0.4pt}}
 \sum_{q} 
 \sum_{X_A,X_B} \! e_q^2 \,
 (2 \pi)^4 \, \delta^4 (p_A-p_a-p_{AX}) \,
 (2 \pi)^4 \, \delta^4 (p_B-p_b-p_{BX}) 
\nonumber \\[-0.10cm]
& 
\times
\left [ \, 
        \langle X_B \left | \bar \psi_{b,j} (0) \right | p_B s_B \rangle \,
             (\Gamma_{q\bar q \to \gamma^* g, \, \mu})_{ji} \,
        \langle X_A \left | \psi_{a,i} (0) \right | p_A s_A \rangle 
             \, \right ] ^\dagger
\left [ \, 
 \langle X_B \left | \bar \psi_{b,l} (0) \right | p_B s_B \rangle \,
             (\Gamma_{q\bar q \to \gamma^* g, \, \nu})_{lk} \,
           \langle X_A \left | \psi_{a,k} (0) \right | p_A s_A \rangle        
             \, \right ] 
\nonumber \\[-0.00cm]
& 
\times
 (2 \pi)^4 \, \delta^4 (p_a+p_b-q-p_d) \, \frac{d^3 p_d}{2E_d \, (2\pi)^3} ,
\label{eqn:cross-lepton-hadron-tensors}
\end{align}
for the $q\bar q \to\gamma^* g$ process.
\end{widetext}

We try to write the hadron tensor in terms of the correlation functions.
The $\delta$ functions $\delta^4 (p_H-p_h-p_{H_X})$ 
($H=A$ or $B$, $h=a$ or $b$) 
in Eq\,(\ref{eqn:cross-lepton-hadron-tensors})
are expressed by the integrals of exponential functions:
$(2 \pi)^4 \, \delta^4 (p_H-p_h-p_{H_X}) 
    = \int d^4 \xi_h e^{-i(p_H-p_h-p_{H_X}) \cdot \xi_h}$.
Then, the quark field is given at $\xi_h$ in the matrix elements
with the exponential factor as
$ e^{i p_{H_X} \cdot \xi_h} \psi (0) e^{-ip_H \cdot \xi_h} = \psi (\xi_h)$.
The completeness relation $\sum_{X_H} \left | X_H \rangle \langle X_H \right | =1$
is used to express the hadron tensor in terms of parton correlation functions 
$\Phi_{h/H} (p_h, p_H, s_H)$, 
which include parton-spin summations and averages, as
\begin{align}
W_{\mu\nu} (q\bar q) & = 2\pi 
  \sum_{\lambda_d}
  \sum_{\text{color}}^{\rule{0.5cm}{0.4pt}}
  \sum_{q} e_q^2
  \int d^4 p_a \int d^4 p_b \, \delta(p_d^2)
\nonumber \\[-0.10cm]
& \ \ \ 
\times 
\text{Tr} 
\left [ \Gamma_{q\bar q \to \gamma^* g, \, \nu} \, \Phi_{q/A} (p_a, p_A, s_A) \, 
\right.
\nonumber \\[-0.10cm]
& \ \ \ \ \ \ \ \,
\times \left.
\hat \Gamma_{q\bar q \to \gamma^* g, \, \mu}
\, \Phi_{\bar q/B} (p_b, p_B, s_B) 
\right ] 
.
\label{eqn:hadron-tensor}
\end{align}
Here, the summation is taken over the gluon spin $\lambda_d$.
In this way, the cross section for $p+d \to \gamma^* +X$ is generally
expressed by the parton correlation functions defined by
the matrix elements of the bilocal operators
in Eqs.\,(\ref{eqn:correlation-q}) and (\ref{eqn:correlation-qbar}).

Using the integrated collinear correlation functions 
with the lightcone relations $d^4 p_h = dp_h^+ dp_h^- d^2 p_{hT}$ 
and $p_h ^\pm = x_h p_H^\pm$,
and using the integral $\int dx_b \delta (p_d^2)$ of Eq.\,(\ref{eqn:pd2}),
we have the hadron tensor:
\begin{align}
& 
W_{\mu\nu} (q\bar q) =  
  \sum_{\lambda_d}
 \sum_{\text{color}}^{\rule{0.5cm}{0.4pt}}
 \sum_{q} e_q^2 
 \int_{\text{min}(x_a)}^1 dx_a \, \frac{\pi}{x_a-x_1}
\nonumber \\[-0.00cm]
& \ \,
\times 
\text{Tr} \left [ \, \Gamma_{q\bar q \to \gamma^* g, \, \nu}
 \, \Phi_{q/A} (x_a) \,
\hat \Gamma_{q\bar q \to \gamma^* g, \, \mu} 
\, \Phi_{\bar q/B} (x_b) \, \right ] .
\label{eqn:hadron-tensor-2}
\end{align}
The momentum fraction $x_b$ for a parton in the deuteron
is defined in Eq.\,(\ref{eqn:p-ab}), so its upper bound is one.
However, one should be careful about this kinematical region.
In lepton DIS, the Bjorken scaling variable
for the nucleon $x_{Bj} = Q^2 /(2 M_N \nu)$ is usually used also
for the deuteron. In this case, due to the difference
between the nucleon and deuteron masses ($M_d/M_N \simeq 2$),
the upper bound is $\max (x_{Bj}) \simeq 2$.
One should note this convention difference in numerical estimates.
If we provide the correlation functions 
$\Phi_{q/A} (x_a)$,  $\Phi_{\bar q/B} (x_b)$ and 
the parton-interaction part 
$\Gamma_{q\bar q \to \gamma^* g, \, \mu}$
in Eq.\,(\ref{eqn:hadron-tensor-2}), 
the Drell-Yan cross section can be evaluated.
We explained relevant PDFs of the proton and deuteron 
in connection with their parton correlation functions
in Sec.\,\ref{pd-pdfs}
for evaluating the partonic cross section in Sec.\,\ref{parton-process}.
From the hadron tensor of Eq.\,(\ref{eqn:hadron-tensor-2})
with the unpolarized correlation functions
$\Phi_{q/A} (x_a) = \slashed{\bar n} f_{q/A} (x_a) /2$
and
$\Phi_{\bar q/A} (x_b) = \slashed{n} f_{\bar q/B} (x_b) /2$
with $\slashed{\bar n} = \gamma^-$ and
$\slashed{n} = \gamma^+ $,
the cross section of Eq.\,(\ref{eqn:cross-parton-I}) becomes
identical to Eq.\,(\ref{eqn:cross-parton-II}).

\begin{figure}[b]
\vspace{-0.00cm}
\begin{center}
   \includegraphics[width=8.0cm]{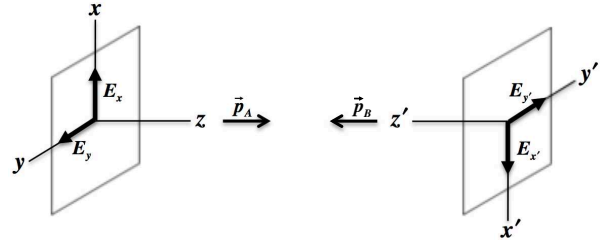}
\end{center}
\vspace{-0.6cm}
\caption{Linear polarizations $E_{x'}$ and $E_{y'}$ 
of spin-1 deuteron in proton($A$)-deuteron($B$) collision.}
\label{fig:linear-pol-B}
\vspace{-0.0cm}
\end{figure}

Before stepping into the cross-section calculation with the gluon transversity,
we comment on the factorization of the cross section into the soft-physics part 
of the correlation functions and the hard part described by perturbative QCD.
As it is explained in Ref.\,\cite{QS-fact}, the Drell-Yan 
cross sections are factorized in the leading order (LO) of $\alpha_s$ 
if the hard scale $Q^2$ is reasonably large. 
The factorization is satisfied in the reasonably
large $Q^2$ region, so that it is theoretically meaningful
to extract the gluon transversity distribution
from experimental measurements.

\begin{figure}[t]
\vspace{-0.00cm}
\begin{center}
   \includegraphics[width=6.5cm]{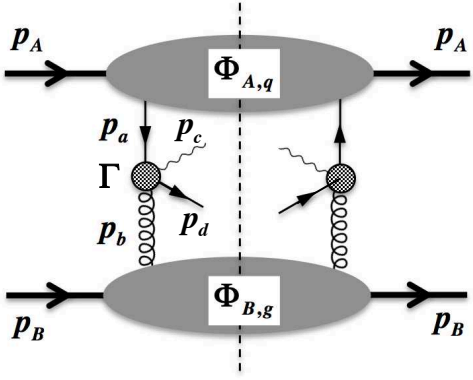}
\end{center}
\vspace{-0.6cm}
\caption{Quark-gluon process contribution to the cross section.}
\label{fig:pd-parton-qg}
\vspace{-0.0cm}
\end{figure}

In this work, the linear polarization asymmetry $E_x -E_y$, 
is taken for the deuteron as illustrated in Fig.\,\ref{fig:linear-pol-B}, 
so that only the $h_{1TT,g/B}(x)$ ($\equiv - \Delta_T g _B (x)$ in this work)
term of Eq.\,(\ref{eqn:correlation-integrated-g}) contributes.
Actually, the asymmetry is $E_{x'} -E_{y'}$ because the deuteron
moves in the negative-$z$ direction in the c.m. frame
as shown in Fig.\,\ref{fig:linear-pol-B}.
However, according to Eq.\,(\ref{eqn:gluon-polarization-tensor-1})
by changing the gluon polarization $\varepsilon$ 
for the deuteron one $E$, they are same: 
$E_{x'} -E_{y'} = E_{-x} -E_{-y} = E_{x} -E_{y}$.
Namely, we leave only the $S_{B,TT}^{\alpha\beta}$ term
and terminate other unpolarized and spin-dependent ones, 
as explained in the end of Sec.\,\ref{d-polarizations},
in the correlation function of 
Eq.\,(\ref{eqn:correlation-integrated-g}).
In the collinear formalism, the hadron tensor is given 
in the same way with  Eq.\,(\ref{eqn:hadron-tensor-2}) by
\begin{align}
& \! \! \!
W_{\mu\nu} (E_{x} -E_{y}) = 
  \sum_{\lambda_d}
 \sum_{\text{color}}^{\rule{0.5cm}{0.4pt}}
 \sum_{q}  e_q^2 
 \int _{\text{min}(x_a)}^1 dx_a \,
 \frac{\pi}{p_g^- (x_a - x_1)}
\nonumber \\[-0.00cm]
& 
\times  \text{Tr}
 \left [ \, \Gamma_{\nu\beta} \!
 \left \{ \Phi_{q/A} (x_a) + \Phi_{\bar q/A} (x_a) \right \}  \!
 \hat\Gamma_{\mu\alpha} 
 \, \Phi_{g/B}^{\,\alpha\beta} (x_b) \, \right ] ,
\label{eqn:hadron-tensor-3}
\end{align}
as illustrated in Fig.\,\ref{fig:pd-parton-qg}.
The summation is taken over the quark spin $\lambda_d$.
Here, only the linear polarization is considered
for the deuteron
by taking the $S_{B,TT}^{\,\alpha\beta}$ term
as the only one to a finite spin asymmetry.
Then, the only contribution comes from the gluon transversity 
$\Delta_T g _B (x)$ 
in the deuteron.
There is no such polarization term 
in the quark and antiquark distributions of the deuteron according to 
Eq.\,(\ref{eqn:correlation-integrated-q-deuteron}).
In order to have the virtual photon in the intermediate stage, 
a charged parton, namely a quark or an antiquark,
needs to be involved in the reaction, 
so that only quark and antiquark correlation functions contribute
as the leading process from the proton.
The collinear correlation functions in 
Eqs.\,(\ref{eqn:correlation-integrated-q-proton}) and 
(\ref{eqn:correlation-integrated-g}) are used
for calculating the hadron tensor of Eq.\,(\ref{eqn:hadron-tensor-3}).
Then, we find that the only contribution from the proton part should 
be the twist-2 PDF term $f_{1,q/A}$.

In this case of polarization, the trace of 
Eq.\,(\ref{eqn:hadron-tensor-3}) typically looks like
\begin{align}
\text{Tr} & \left [ \slashed{p}_d \, \gamma_\beta \, 
\slashed{p}_i \, \gamma_\nu \,  
\left \{ \Phi_{q/A} (x_a) + \Phi_{\bar q/A} (x_a) \right \}
\cdots 
\right.
\nonumber \\
& \ \ 
\times
\left.
\gamma_\mu \, \slashed{p}_i \, \gamma_\alpha \, \, 
\Phi_{g/B}^{\, \alpha\beta} (x_b) \right ] ,
\label{eqn:typical-trace}
\end{align}
where 
$\gamma_\alpha$ and $\gamma_\beta$ are from the quark-gluon vertex,
and $p_i$ is the intermediate-quark momentum, for example, $p_i =p_a + p_b$.
There could be a factor $\slashed{p}_a$, 
for example, in an unpolarized-proton reaction.
However, it is included into the definition of $\Phi_{q/A}$,
and it corresponds to the first $\slashed{\bar n}$
in Eq.\,(\ref{eqn:correlation-integrated-q-proton}).
There are seven $\gamma$ matrices, except for the ones in $\Phi_{q(\bar q)/A}$, 
in the trace and there is no $\gamma$ factor 
in the gluon transversity term 
$ \Delta_T g_B (x_b)$ of $\Phi_{g/B}$, 
so that only odd numbers of $\gamma$ in $\Phi_{q/A} (x)$ 
contribute to the spin asymmetry. 
In addition, the unpolarized-proton beam is assumed
in this work, so that only the unpolarized distribution 
$\slashed{\bar n} \, f_{1,q/A} (x_a)/2$
contributes to the hadron tensor.

In this way, we find that the leading contribution starts 
from the twist-2 distribution $f_{1,q/A}$ in the nucleon part
by considering that only the gluon transversity $\Delta_T g_B$ participates 
from the deuteron for the linear-polarization asymmetry.
From Eqs.\,(\ref{eqn:correlation-integrated-q-proton}) 
and (\ref{eqn:correlation-integrated-g}), they are given by
\begin{align}
\Phi_{q/A} (x_a) 
& = \frac{1}{2} \, \slashed{\bar n} \, q_A (x_a) ,
\nonumber \\
\Phi_{g/B}^{\, \alpha\beta} (x_b) 
& =  - S_{B,TT}^{\, \alpha\beta} \, \Delta_T g_B (x_b) .
\label{eqn:contribute-pdfs}
\end{align}
Here, $\Phi_{g/B}^{\, \alpha\beta} (x_b)$ is multiplied
by the factor of 2 because the linear polarization
$E_{x} -E_{y}$ is taken for showing the cross section instead of
$(E_{x} -E_{y})/2$, and $f_{1,q/A} (x_a)$ is denoted as $q_A (x_a)$.

\subsection{Parton-interaction processes}
\label{parton-process}

\begin{figure}[b]
\vspace{-0.00cm}
\begin{center}
   \includegraphics[width=8.5cm]{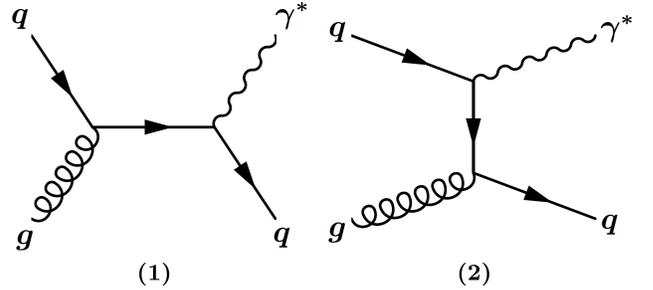}
\end{center}
\vspace{-0.6cm}
\caption
{Leading partonic processes $q (p_a) + g (p_b) \to \gamma^* (p_c=q) + q (p_d)$:
 $(1)$ $s$-channel and $(2)$ $t$-channel processes.}
\label{fig:parton-process}
\vspace{-0.00cm}
\end{figure}

For calculating the hadron tensor of Eq.\,(\ref{eqn:hadron-tensor-3})
and subsequently the cross section of Eq.\,(\ref{eqn:cross-parton-I}),
partonic matrix elements are calculated in this section. 
We consider the leading partonic processes 
in Fig.\,\ref{fig:parton-process}.
Since the linear polarization spin asymmetry in 
Eq.\,(\ref{eqn:polarized-cross-1}) is studied in this work
and since unpolarized and polarized gluon distributions in the proton
do not contribute as the leading process,
main contributions come from the partonic reactions of 
\begin{equation}
q\,/\,\bar q \text{ in proton}+g \text{ in deuteron}
\to \gamma^* + q 
\nonumber
\label{eqn:leading-process}
\end{equation}
as shown in Fig.\,\ref{fig:parton-process}.

The parton-scattering amplitude of Fig.\,\ref{fig:parton-process}
is given by
\begin{align}
\Gamma _{\mu\alpha} & =  i \, g \, ( t^{a} )_{ji} 
                           \, \bar u (p_d) \, G_{\mu\alpha},
\nonumber \\
& 
G_{\mu\alpha} \equiv 
  \gamma_{\mu} \frac{\slashed{p}_a+ \slashed{p}_b}{\hat s} \gamma_{\alpha}
 + \gamma_{\alpha} \frac{\slashed{p}_a- \slashed{q}}{\hat t} \gamma_{\mu}  ,
\label{eqn:parton-matrix-1}
\end{align}
where $g$ is the strong coupling constant.
Then, the color factor in $W_{\mu\nu}$ becomes
\begin{align}
\sum_{\text{color}}^{\rule{0.5cm}{0.4pt}}
 (t^{a})_{ji} (t^{a})_{ji}^\dagger
& = \frac{1}{3 \cdot 8} \sum_{a} (t^{a})_{ji} (t^{a})_{ji}^\dagger
= \frac{1}{8} \, C_F , 
\nonumber \\
C_F & = \frac{N_c^2 -1}{2 N_c}, \ \ N_c=3 .
\label{eqn:color-factor-1}
\end{align}
Using these expressions, the running coupling constant given 
by $\alpha_s = g^2 /(4 \pi)$, $p_g^-=x_b p_B^-$, 
and the correlation functions in Eq.\,(\ref{eqn:contribute-pdfs}),
we write the hadron tensor of Eq.\,(\ref{eqn:hadron-tensor-3})
in terms of the PDFs as
\begin{align}
& \! \! \!
W_{\mu\nu} (E_{x} -E_{y}) 
 = - \frac{\pi^2 \alpha_s C_F}{4 p_B^- }
 \sum_{q}  e_q^2 
\nonumber \\[-0.00cm]
& \ \ \ \ \ 
\times 
 \int _{\text{min}(x_a)}^1 dx_a \,
 \frac{1}{x_b (x_a - x_1)}
\left\{ \, q_A (x_a) + \bar q_A (x_a) \, \right\}
\nonumber \\[-0.00cm]
& \ \ \ \ \ \ \ \ \ \ \ \ 
\times 
 \Delta_T g_B (x_b) \,
S_{B,TT}^{\, \alpha\beta}
\, \text{Tr}
 \left [ \, \slashed{p}_d \, G_{\nu\beta}  \slashed{\bar n} \,
 \hat G_{\mu\alpha}  \, \right ] .
\label{eqn:hadron-tensor-5}
\end{align}

Next, the trace should be calculated. 
We need trace calculations with eight $\gamma$ matrices and it is 
slightly lengthy, so that their results should be carefully
checked. Three independent methods are used.
First, we analytically calculated the trace by noting
the $\mu\nu$ and $\alpha\beta$ symmetries of $g^{\mu \nu}$ 
and $S_{B,TT}^{\, \alpha\beta}$ together with 
properties of spin and momentum vectors within the trace.
As the independent second method, it is calculated by using 
the FeynCalc code \cite{feyncalc} together with Mathematica.
The third one is by using the Tracer code \cite{tracer}
with also Mathematica. 
All these results are consistent with each other.

We divide the trace into $s$-channel, $t$-channel,
and their interference terms:
\begin{align}
& 
g^{\mu \nu} \, S_{B,TT}^{\, \alpha\beta} \, 
\text{Tr}
 \left [ \, \slashed{p}_d \, G_{\nu\beta} \, \slashed{\bar n} \,
 \hat G_{\mu\alpha}  \, \right ]
\nonumber \\
& 
= \frac{1}{\hat s^2}   | M_s |^2 
\! + \! \frac{1}{\hat t^2} | M_t |^2 
\! + \! \frac{1}{\hat s \, \hat t} 
     ( M_s^* M_t + M_s M_t^* )  .
\label{align:trace-terms}
\end{align}
Each term is calculated as
\begin{align}
| M_s |^2 
& = g^{\mu \nu} \, S_{B,TT}^{\, \alpha\beta} \, 
\text{Tr} \,
[ \slashed{\bar n} \gamma_\alpha
 (\slashed{p}_a + \slashed{p}_b ) \gamma_\mu \, \slashed{p}_d 
 \gamma_\nu  (\slashed{p}_a + \slashed{p}_b ) \gamma_\beta ]
\nonumber \\
& 
= 16 \, \hat s \, \bar n^\zeta S_{B,TT,\zeta}^\alpha \, p_{d,\alpha} =0,
\label{align:trace-s}
\end{align}
\begin{align}
| M_t |^2
&  = g^{\mu \nu} \, S_{B,TT}^{\, \alpha\beta} \, 
\text{Tr} \,
[ \slashed{\bar n} \gamma_\mu 
 (\slashed{p}_a - \slashed{p}_c ) \gamma_\alpha \, \slashed{p}_d 
 \gamma_\beta  (\slashed{p}_a - \slashed{p}_c ) \gamma_\nu ]
\nonumber \\
& 
= -32 (q_\alpha S_{B,TT}^{\, \alpha\beta} q_\beta ) 
   \, \bar n \cdot (p_a-p_c) ,
\label{align:trace-t}
\end{align}
\begin{align}
& \! \! 
M_s^* M_t + M_s M_t^*
\nonumber \\
& =  g^{\mu \nu} \, S_{B,TT}^{\, \alpha\beta} 
\left\{ 
\text{Tr} \,
[ \slashed{\bar n} \gamma_\alpha
 (\slashed{p}_a + \slashed{p}_b ) \gamma_\mu \, \slashed{p}_d 
 \gamma_\beta  (\slashed{p}_a - \slashed{p}_c ) \gamma_\nu ]
 \right.
\nonumber \\
& \ \ \ \ \ \ \ \ \ \ \ \ \ \ \ \
\left.
+ \text{Tr} \,
[ \slashed{\bar n} \gamma_\mu 
 (\slashed{p}_a - \slashed{p}_c ) \gamma_\alpha \, \slashed{p}_d 
 \gamma_\nu  (\slashed{p}_a + \slashed{p}_b ) \gamma_\beta ]
 \right \}
\nonumber \\
& \!
= 32 (q_\alpha S_{B,TT}^{\, \alpha\beta} q_\beta )
   \, \bar n \cdot (p_a+p_b) .
\label{align:trace-st-int}
\end{align}
Noting Eq.\,(\ref{eqn:eps-contraction-1}) and 
the relations 
$\bar n \cdot (p_a+p_b) = \sqrt{s/2} \, x_b$ 
and $\bar n \cdot (p_a-p_c) = - \sqrt{s/2} \, x_2$,
we obtain
\begin{align}
& g^{\mu \nu} \, S_{B,TT}^{\, \alpha\beta} \, 
\text{Tr}
 \left [ \, \slashed{p}_d \, G_{\nu\beta}  \, \slashed{\bar n} \,
 \hat G_{\mu\alpha}  \, \right ]
\nonumber \\
& \ \ \ 
=  - \frac{32 \, Q^2 \, q_T^2}
  {\sqrt{2} \, s^{5/2} \, x_a \,(\tau-x_a x_2)^2} \cos (2\phi) .
\label{eqn:trace-terms}
\end{align}
Substituting Eq.\,(\ref{eqn:trace-terms}) into
Eq.\,(\ref{eqn:cross-parton-I}),
we finally obtain the cross section as
\begin{align}
& 
\frac{ d \sigma_{pd \to \mu^+ \mu^- X} }{d\tau \, dq_T^2 \, d\phi \, dy}
(E_x-E_y )
= - \frac{\alpha^2 \, \alpha_s \, C_F \, q_T^2}{6\pi s^3} \cos (2\phi) 
\nonumber \\
& \ \ \ 
\times
\int_{\text{min}(x_a)}^1 dx_a 
 \frac{1} { (x_a x_b)^2 \, (x_a -x_1) (\tau -x_a x_2 )^2}
\nonumber \\
&  \ \ \ \ \ \ \ 
 \times 
 \sum_{q}  e_q^2 \, x_a \!
 \left[ \, q_A (x_a) + \bar q_A (x_a) \, \right ]
  x_b \Delta_T g_B (x_b) .
\label{eqn:cross-5}
\end{align}
Here, the deuteron polarization is explicitly written as 
$E_x-E_y$, which indicates the polarization asymmetry of 
Eq.\,(\ref{eqn:polarized-cross-1}).

Actual polarization measurements are usually done 
by polarization asymmetries. For such estimations, the cross section
$d \sigma_{pd \to \mu^+ \mu^- X} (E_x+E_y )$ should be calculated
in the same way.
There are two process types, $q \bar q \to \gamma^* g$
and $q \, (\text{or } \bar q) g \to \gamma^* q \, (\text{or } \bar q)$.
For the $q \bar q \to \gamma^* g$ processes,
the hadron tensor, which corresponds to Eq.\,(\ref{eqn:hadron-tensor-2}),
is given by
\begin{align}
& 
\! \! 
W_{\mu\nu} (q\bar q, E_{x} +E_{y}) =  
  \sum_{\lambda_d}
 \sum_{\text{color}}^{\rule{0.5cm}{0.4pt}}
 \sum_{q} e_q^2 
 \int_{\text{min}(x_a)}^1 dx_a \, \frac{\pi}{x_a-x_1}
\nonumber \\[-0.10cm]
& \ \,
\times 
\text{Tr} \bigg [ \, \Gamma_{q\bar q \to \gamma^* g, \, \nu}
 \, \Phi_{q/A} (x_a) \,
\hat \Gamma_{q\bar q \to \gamma^* g, \, \mu} 
\, \Phi_{\bar q/B} (x_b)
\nonumber \\[-0.10cm]
& \ \ \ \ \ \ \ 
+ \Gamma_{\bar q q \to \gamma^* g, \, \nu}
 \, \Phi_{\bar q/A} (x_a) \,
\hat \Gamma_{\bar q q \to \gamma^* g, \, \mu} 
\, \Phi_{q/B} (x_b)
\, \bigg ] .
\label{eqn:hadron-tensor-qqbar}
\end{align}
For the $q \, (\text{or } \bar q) g \to \gamma^* q \, (\text{or } \bar q)$
processes, the hadron tensor, which corresponds to 
Eq.\,(\ref{eqn:hadron-tensor-3}), is given by
\begin{align}
& \! 
W_{\mu\nu} (qg, E_{x} +E_{y}) = 
  \sum_{\lambda_d}
 \sum_{\text{color}}^{\rule{0.5cm}{0.4pt}}
 \sum_{q}  e_q^2 
 \int _{\text{min}(x_a)}^1 dx_a \,
 \frac{\pi}{x_a - x_1}
\nonumber \\[-0.10cm]
& \! \! \! \! \! \!
\times  \text{Tr}
 \bigg [ \, 
 \frac{1}{p_g^-} \Gamma_{qg \to \gamma^* q, \nu\beta} \!
 \left \{ \Phi_{q/A} (x_a) + \Phi_{\bar q/A} (x_a) \right \}  \!
 \hat\Gamma_{qg \to \gamma^* q, \mu\alpha} 
 \, \Phi_{g/B}^{\,\alpha\beta} (x_b) 
\nonumber \\[-0.10cm]
& \! \! \! \! \!
+ \frac{1}{p_g^+} \Gamma_{gq \to \gamma^* q, \nu\beta} \!
 \left \{ \Phi_{q/B} (x_b) + \Phi_{\bar q/B} (x_b) \right \}  \!
 \hat\Gamma_{gq \to \gamma^* q, \mu\alpha} 
 \, \Phi_{g/A}^{\,\alpha\beta} (x_a) 
 \, \bigg ] .
\label{eqn:hadron-tensor-qg}
\end{align}

Here, the linear polarizations $E_x$ and $E_y$ are considered 
for the deuteron. 
For the polarization $E_x$ ($E_y$), the spin vector and tensor
of Eq.\,(\ref{eqn:spin-1-vector-tensor-2})
are given by $\vec S = 0$ and
$T_{ij} = \delta_{ij}/3 -\delta_{i1}\delta_{j1}$ 
($T_{ij} = \delta_{ij}/3 -\delta_{i2}\delta_{j2}$), 
which indicate
$S_{B,T}^x = S_{B,T}^y = S_{B,L} = S_{B,TT}^{xy} = S_{B,LT}^{x} = S_{B,LT}^{y}=0$,
$S_{B,LL} = 1/2$ and $S_{B,TT}^{xx}=-1$ ($S_{B,TT}^{xx}=1$ for $E_y$)
in Eq.\,(\ref{eqn:spin-1-vector-tensor})
for the deuteron ($B$).
Then, the correlation functions of
Eqs.\,(\ref{eqn:correlation-integrated-q-deuteron})
and (\ref{eqn:correlation-integrated-g}) for the deuteron become
\begin{align}
& \Phi_{q/B}  (x_b)_{E_x} + \Phi_{q/B}  (x_b)_{E_y}
\nonumber \\
& \ \ \ 
 = \slashed{n}  \, f_{1,q/B} (x_b) 
  + \slashed{n} \, S_{B,LL} \, f_{1 LL, q/B} (x_b) ,
\nonumber \\
&  
\Phi_{g/B}^{\, \alpha\beta} (x_b)_{E_x} + \Phi_{g/B}^{\, \alpha\beta} (x_b)_{E_y}
\nonumber \\
& \ \ \ 
=  - g_T^{\, \alpha\beta} \, f_{1,g/B} (x_b) 
- g_T^{\, \alpha\beta} \, S_{B,LL} \, f_{1LL,g/B} (x_b) .
\label{eqn:correlation-integrated-qg-2}
\end{align}
Therefore, the collinear quark and gluon correlation functions are expressed
by the unpolarized PDFs $f_{1,q/B} (x_b)$ and $f_{1,g/B} (x_b)$
and the tensor-polarized PDFs $f_{1 LL, q/B} (x_b)$ and $f_{1LL,g/B} (x_b)$.
Although there is some information of the tensor-polarized quark distributions
$f_{1 LL, q/B} (x_b)$ from the HERMES measurement \cite{jlab-b1,b1-convolution},
there are no established distributions at this stage. 
A finite tensor-polarized gluon distribution should appear through the $Q^2$
evolution \cite{fermilab-pd}; however, there is no reliable distribution at this stage.
In any case, the tensor-polarized PDFs are of the order of a few percent
or less in comparison with the unpolarized PDFs 
\cite{jlab-b1,b1-convolution,fermilab-pd},
they are neglected in numerical analysis of this work by taking
$ \Phi_{q/B}  (x_b)_{E_x} + \Phi_{q/B}  (x_b)_{E_y} 
  =\slashed{n}  \, f_{1,q/B} (x_b)$ and
$ \Phi_{g/B}^{\, \alpha\beta} (x_b)_{E_x} + \Phi_{g/B}^{\, \alpha\beta} (x_b)_{E_y}
  = - g_T^{\, \alpha\beta} \, f_{1,g/B} (x_b)$.

\begin{widetext}
Repeating similar calculations for the cross section, we obtain 
\begin{align}
& \frac{ d \sigma_{pd \to \mu^+ \mu^- X} }{d\tau \, dq_T^2 \, d\phi \, dy} 
(E_x+E_y )
=  \frac{\alpha^2 \, \alpha_s \, C_F}{2 \pi Q^2} 
\int_{\text{min}(x_a)}^1 dx_a \frac{1}{x_a -x_1}
\sum_{q}  e_q^2 \, \bigg [ \frac{4}{9} 
\left\{ q_A (x_a) \, \bar q_B (x_b) + \bar q_A (x_a) \, q_B (x_b) \right\}
\frac{2Q^2 \hat s + \hat t^2 + \hat u^2}{\hat s \, \hat t \, \hat u}
\nonumber \\
&  \ \ \ \ \ \ \ \ \ \ \ \ \ \ \ 
 - \frac{1}{6 \, \hat s} \, \bigg (
 \left\{ q_A (x_a) + \bar q_A (x_a) \right\} g_B (x_b)
 \frac{2Q^2 \hat u + \hat s^2 + \hat t^2}{\hat s \, \hat t}
 + g_A (x_a) \left\{ q_B (x_b) + \bar q_B (x_b) \right\} 
 \frac{2Q^2 \hat t + \hat s^2 + \hat u^2}{\hat s \, \hat u}
 \bigg ) \bigg ].
\label{eqn:cross-6}
\end{align}
In fact, this expression agrees with the one in Ref.\,\cite{reya-1981}.
There is an overall factor of 2 in Eq.\,(\ref{eqn:cross-6})
because $d\sigma (E_x)+d\sigma (E_x)$ is calculated instead 
of the spin average in Ref.\,\cite{reya-1981}.
The cross section is expressed by the variables $x_a,\, x_b,\, x_1,\, x_2$ as
\begin{align}
& \! 
\frac{ d \sigma_{pd \to \mu^+ \mu^- X} }{d\tau \, dq_T^2 \, d\phi \, dy}
(E_x+E_y )
=  \frac{\alpha^2 \, \alpha_s \, C_F}{2 \pi \, \tau \, s^2} 
\int_{\text{min}(x_a)}^1 dx_a \frac{1}{(x_a -x_1) \, x_a^2 \, x_b^2}
\nonumber \\
& 
\times \sum_{q}  e_q^2 \, \bigg [ \frac{4}{9} 
\left\{ q_A (x_a) \, \bar q_B (x_b) + \bar q_A (x_a) \, q_B (x_b) \right\}
\frac{2 \tau \left\{\tau -(-2 x_a x_b +x_1 x_b +x_2 x_a) \right\} 
                    +x_b^2 (x_a-x_1)^2+x_a^2 (x_b-x_2)^2 }{(x_a-x_1)(x_b-x_2)}
\nonumber \\
& \ \ \ \ \ \ \ \ \ \ \ \ 
 + \frac{1}{6} \left\{ q_A (x_a) + \bar q_A (x_a) \right\} g_B (x_b)
     \frac{ 2\tau (\tau-x_1 x_b) +x_b^2 \left\{ (x_a-x_1)^2 + x_a^2 \right\}}
          {x_b (x_a-x_1)}
\nonumber \\
& \ \ \ \ \ \ \ \ \ \ \ \ 
 + \frac{1}{6} \, g_A (x_a) \left\{ q_B (x_b) + \bar q_B (x_b) \right\} 
    \frac{ 2\tau (\tau-x_2 x_a) + x_a^2 \left\{ (x_b-x_2)^2 + x_b^2 \right\}}
         {x_a (x_b-x_2)}
 \bigg ].
\label{eqn:cross-7}
\end{align}
\end{widetext}
Then, the polarization asymmetry 
\begin{align}
A_{E_{xy}} = 
\frac{d \sigma_{pd \to \mu^+ \mu^- X} (E_x-E_y) / (d\tau \, dq_T^2 \, d\phi \, dy)}
     {d \sigma_{pd \to \mu^+ \mu^- X} (E_x+E_y) / (d\tau \, dq_T^2 \, d\phi \, dy)} ,
\label{eqn:pol-asym}
\end{align}
is estimated numerically in Sec.\,\ref{results}.

\section{Numerical Results}
\label{results}

By the formula in Eq.\,(\ref{eqn:cross-5}), the cross section 
is evaluated numerically. We need two types of parton distributions 
for this calculation.
One is the unpolarized PDFs 
$q_A$ and $\bar q_A$
in the proton, and the other is the gluon transversity $\Delta_T g_B$
in the deuteron. Since the unpolarized PDFs are accurately determined
except for extreme kinematical regions,
if one of any recent parametrizations is taken, 
it provides a reasonable estimate on the proton part.
As one of such a parametrization, the LO set of the CTEQ14
\cite{cteq14} is used for our numerical calculations
in the leading-order and leading-twist level. 
There is a charge squared factor in Eq.\,(\ref{eqn:cross-5}),
so that the proton part simply becomes the structure function
$F_2$ in the LO.

Next, an appropriate gluon transversity distribution should be
taken. However, due to the lack of experimental information, 
there is no realistic gluon transversity distribution 
for evaluating the cross section. 
Since it is the purpose of this work to propose a possible experiment 
to find it at hadron accelerator facilities, we need to have a rough
estimate on the magnitude of the cross section 
for future experimental proposals.
We note that the quark transversity distributions in the proton
are expected to be equal to the longitudinally-polarized quark
distributions at small $Q^2$, where the longitudinal polarization
and the transverse one do not matter. From this consideration,
we may boldly assume at first that the gluon transversity distribution 
is the same as the longitudinally-polarized gluon distribution
for calculating the cross section, just as a rough order 
of magnitude estimate on the cross section. 
However, we should note that the longitudinally-polarized 
gluon distribution and the gluon transversity distribution
have different physics origins, so that 
the actual cross section would be different.
Of course, such an assumed distribution should be actually
measured by future experiments.
Later, we show how the cross section depends on this assumption.

\begin{figure}[t]
\vspace{-0.00cm}
\begin{center}
   \includegraphics[width=6.0cm]{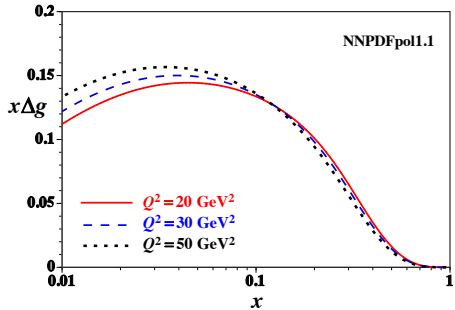}
\end{center}
\vspace{-0.6cm}
\caption{Used longitudinally-polarized gluon distributions
of the NNPDFpol1.1 parametrization
are shown at $Q^2=20$, $30$, and $50$ GeV$^2$.
These gluon distributions are used for a rough estimate on
the Drell-Yan cross sections.
}
\label{fig:nnpdf-gluon}
\vspace{-0.00cm}
\end{figure}

The longitudinally-polarized gluon distribution is not well 
determined at this stage, and it is one of the major purposes 
of building the future electron-ion collider. 
We take one of recent parametrizations, obtained by global analysis
of world data in polarized proton reactions, 
on the longitudinally-polarized gluon distribution.
The employed set is the Neural Network (NN) PDFpol1.1 
version \cite{nnpdfpol1.1}.
The NNPDFpol gluon distributions are shown in Fig.\,\ref{fig:nnpdf-gluon}
at $Q^2=20$, $30$, and $50$ GeV$^2$.
Since the dimuon cross sections are measured in the Fermilab
Drell-Yan experiment in the dimuon-mass region,
$4^2 < M_{\mu\mu}^2 = Q^2 < 9^2$ GeV$^2$, we consider
the scale dependence in this range.
The NNPDF determination reflects the 
Relativistic Heavy Ion Collider (RHIC) pion- and jet-production
measurements, which are sensitive to the polarized gluon distribution,
and it is one of reliable models at this stage. 

There are two factors which need to be carefully considered
in using the nucleonic PDFs for the deuteron ones.
The first point is to assume that the deuteron PDFs 
are simply given by the addition of proton and neutron
contributions as the first approximation.
In our case, the gluon transversity of the deuteron
is tentatively assumed as the addition of the longitudinally-polarized
gluon distributions for the proton and neutron:
$\Delta_T g_d = \Delta g_p + \Delta g_n$.

The second point is that one needs to be careful 
about the scaling variable or the momentum fraction,
as discussed after Eq.\,(\ref{eqn:hadron-tensor-2}).
In the Fermilab Drell-Yan experiment, the momentum fraction 
for a parton $q$ in the deuteron
is defined by $p_q = x_2 (p_d /2)$, namely by using the deuteron momentum
per nucleon, in Fermilab Drell-Yan experiments \cite{wen-chen-discuss}, 
so that its kinematical range is, in principle, $0 \le x_2 \leq 2$. 
However, the range $1 \le x_2 \leq 2$ is usually neglected and
it is not even shown because the PDFs are tiny and it cannot be reached 
by experiments \cite{Fermilab-dy,wen-chen-discuss}. 
This definition of $x_2$ corresponds to 
the Bjorken scaling variable $x = Q^2 /(2 M_N \nu)$ ($0 \le x \leq 2$)
used in lepton DIS experiments with the deuteron target.

The momentum fraction $x_b$ has been used in this article
for partons in the deuteron, and its range is given by
$0 \leq x_b \leq 1$ by definition. 
If deuteron structure functions are assumed to be
a simple addition of proton and neutron ones, 
finite PDFs exist only in the kinematical region
of $0 \leq x_b \leq 1/2$ and they vanish
in the region $1/2 \leq x_b \leq 1$. 
On the other hand, the nucleonic PDFs are provided
in the Bjorken-$x$ region of $0 \leq x_{Bj} \leq 1$,
and the deuteron PDFs exist in the range $0 \leq x_{Bj} \leq 2$.
Since the cross section is formulated by using the variable $x_b$,
we need to express $\Delta_T g_d (x_b)$ in Eq.\,(\ref{eqn:cross-5})
in terms of $\Delta g_{p,n} (x_{Bj})$ for the proton and neutron. 
To preserve the parton densities in changing the scaling variable,
we need to take 
$\Delta_T g_d (x_b) = 2 \, [\Delta g_{p} (x_{Bj}) + \Delta g_{n} (x_{Bj})]$,
where $0 \leq x_{Bj} \leq 1$ for the nucleon,
in evaluation of the cross section. For calculating this relation
in the cross section, we first define the momentum fraction $x_2$, 
in the range $0 \leq x_2 \leq 2$, by $x_2 \equiv 2 x_b$ from $x_b$
which is given by $x_a$ in Eq.\,(\ref{eqn:min-xa}).
Then, the only range $0 \leq x_2 \leq 1$ is used as the scaling variable $x_{Bj}$
for the nucleons.
Alternatively, one can formulate the Drell-Yan cross section
from the beginning by the independent addition of proton-proton 
and proton-neutron cross sections 
\cite{proton-neutron}
to check this kinematical factor of 2.
We actually confirmed such a factor of 2, for example, by considering 
the simple subprocess $q\bar q \to \gamma^* \to \mu^+ \mu^-$ 
for the unpolarized Drell-Yan process.
In fact, such a formalism has been used in the proposals 
of the Fermilab Drell-Yan experiments with 
the definition of the c.m. energy of Eq.\,(\ref{eqn:e906}).

\begin{figure}[t]
\vspace{-0.00cm}
\begin{center}
   \includegraphics[width=8.0cm]{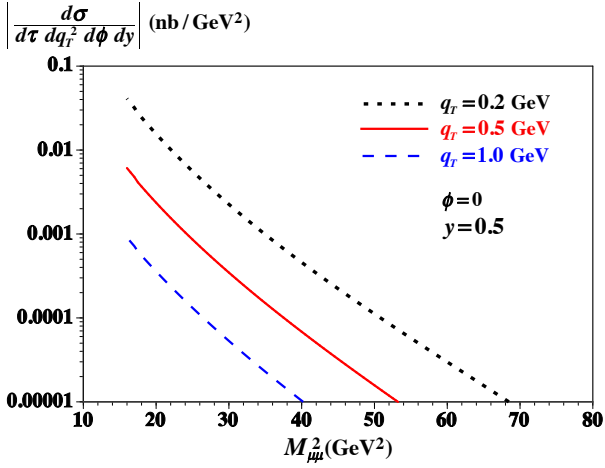}
\end{center}
\vspace{-0.6cm}
\caption{The proton-deuteron Drell-Yan cross sections
are shown as the function of the dimuon-mass squared $M_{\mu\mu}^{\,2}$ at
$q_T=0.2,\, 0.5$, and $1$ GeV for the dimuon azimuthal angle $\phi=0$
and the dimuon rapidity $y=0.5$. The CTEQ14 PDFs are used 
for the unpolarized PDFs of the proton,
and the longitudinally-polarized gluon distribution is taken 
from the NNPDF1.1, and the gluon transversity is assumed 
to be the same as the longitudinal one for numerical estimates.}
\label{fig:dimuon-mass-cross}
\vspace{-0.00cm}
\end{figure}

\begin{figure}[t]
\vspace{-0.00cm}
\begin{center}
   \includegraphics[width=8.0cm]{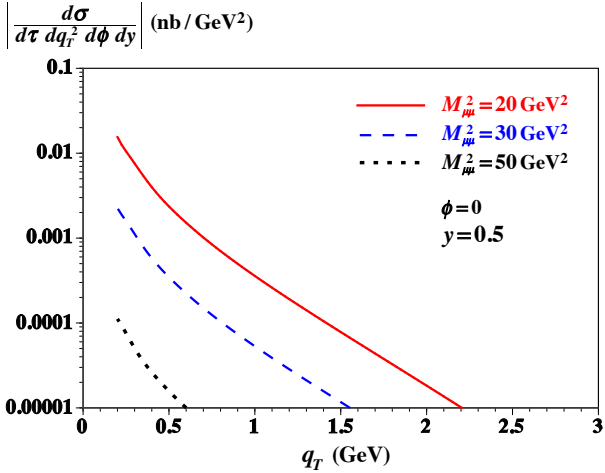}
\end{center}
\vspace{-0.6cm}
\caption{The proton-deuteron Drell-Yan cross sections
are shown as the function of the dimuon transverse momentum $q_T$ 
at the dimuon-mass squared $M_{\mu\mu}^{\,2} =20$, 30, and 50 GeV$^2$
for the azimuthal angle $\phi=0$ and the rapidity $y=0.5$.}
\label{fig:dimuon-pt-cross}
\vspace{-0.30cm}
\end{figure}

\begin{figure}[t]
\vspace{-0.00cm}
\begin{center}
   \includegraphics[width=8.0cm]{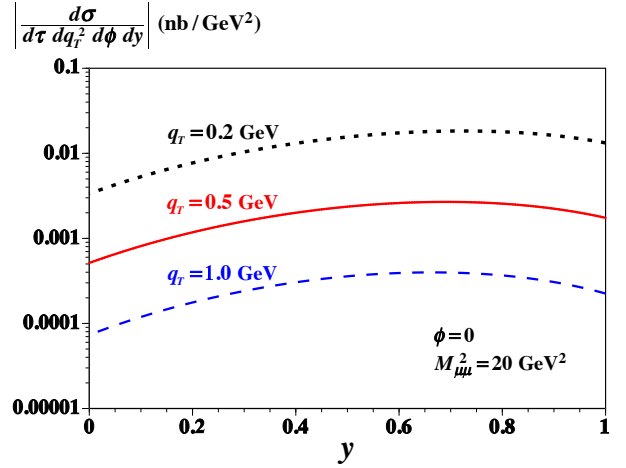}
\end{center}
\vspace{-0.6cm}
\caption{The proton-deuteron Drell-Yan cross sections
are shown as the function of the dimuon rapidity $y$ 
at the dimuon transverse momentum $q_T=0.2$, 0.5, and 1 GeV,
for the azimuthal angle $\phi=0$
and the dimuon-mass squared $M_{\mu\mu}^{\,2} =20$ GeV$^2$.}
\label{fig:dimuon-y-cross}
\vspace{-0.30cm}
\end{figure}

Finally, we show the cross sections obtained by using 
the CTEQ14 for the proton PDFs and the gluon transversity distribution, 
which is assumed to be the longitudinal one in Fig.\,\ref{fig:nnpdf-gluon},
as the function of the dimuon-mass squared ($M_{\mu\mu}^2 = Q^2$) 
in Fig.\,\ref{fig:dimuon-mass-cross}.
The dimuon azimuthal angle and the dimuon rapidity are fixed at 
$\phi=0$ and $y=0.5$. The dotted, solid, and dashed curves
indicate the cross sections at $q_T=0.2$ GeV, 0.5 GeV, and 1.0 GeV,
respectively.
The hard scale $Q^2$ for calculating the PDFs is taken as 
the dimuon-mass squared $Q^2 =M_{\mu\mu}^{\, 2}$. 
The magnitude of the cross section is typically 0.001 $\sim$ 0.1 nb/GeV$^2$
in this kinematical range. 
The dependence of the cross section on the dimuon transverse-momentum
$q_T$ is shown in Fig.\,\ref{fig:dimuon-pt-cross}.
The solid, dashed, and dotted curves indicate the cross sections 
at $M_{\mu\mu}^2=20$, 30, and 50 GeV$^2$, respectively.
In Fig.\,\ref{fig:dimuon-y-cross}, the dependence is shown
on the dimuon rapidity at $q_T=0.2$, 0.5, and 1.0 GeV
by fixing the angle $\phi=0$ and the dimuon-mass squared 
$M_{\mu\mu}^{\,2} =20$ GeV$^2$.

In Figs.\,\ref{fig:dimuon-mass-cross} and \ref{fig:dimuon-pt-cross},
the cross section drops fast as $p_T$ and $M_{\mu\mu}^{\,2} (=Q^2)$ increase.
The kinematical factor in the cross-section integrand 
of Eq.\,(\ref{eqn:cross-5}) and the integral minimum for $x_a$ are given by
\begin{align}
& \! 
\frac{q_T^2} { x_a x_b \, (x_a- x_1) (\tau-x_a x_2)^2 } 
 = \frac{q_T^2}
    {x_a \left ( \frac{Q^2}{s}  -x_a \sqrt{\frac{Q^2 + q_T^2}{s}}e^{-y} \right )^3} ,
\nonumber \\
& \!
\text{min}(x_a)  = \frac{x_1-\tau}{1-x_2} 
   = \frac{\sqrt{\frac{Q^2 + q_T^2}{s}}e^y - \frac{Q^2}{s}}
         {1-\sqrt{\frac{Q^2 + q_T^2}{s}}e^{-y}} .
\label{eqn:kinematical-int}
\end{align}
The first equation indicates that the cross section decreases 
with $Q^2$ and $q_T$. In addition, the minimum of $x_a$
increases with $Q^2$ and $q_T$ at positive rapidity $y$, 
which restricts the integral region.
Since there is a factor $1/(x_a)^2$ in the integrand,
the increase of $Q^2$ and $q_T$ significantly reduces the cross section.
Third, the scale $Q^2$ is given by $M_{\mu\mu}^{\,2}$ and 
the PDFs, $q_A (x_a)$, $\bar q_A (x_a)$, and $\Delta_T g_B(x_b)$, 
change with $Q^2$.
Furthermore, the running coupling constant $\alpha_s$ 
also becomes slightly smaller by the increase of $Q^2$.

\begin{figure}[t]
\vspace{-0.30cm}
\begin{center}
   \includegraphics[width=8.0cm]{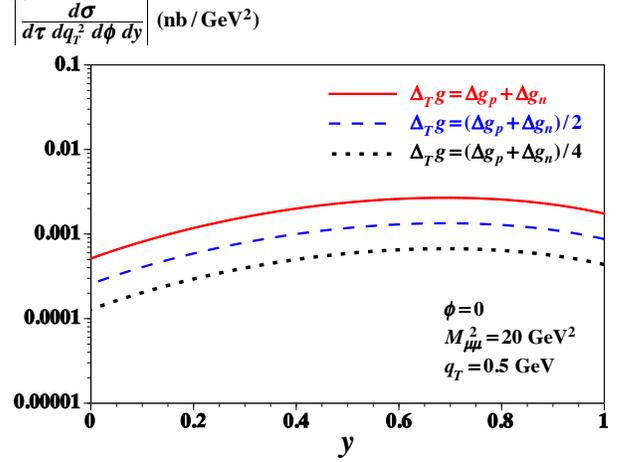}
\end{center}
\vspace{-0.6cm}
\caption{Dependence is shown on the choice of the gluon 
transversity for the proton-deuteron Drell-Yan cross section 
as the function of the rapidity $y$
at the dimuon transverse momentum $q_T=0.5$ GeV
for the azimuthal angle $\phi=0$
and the dimuon-mass squared $M_{\mu\mu}^2 =20$ GeV$^2$.
The NNPDF1.1 gluon distribution is used.
Three gluon transversity distributions are assumed as
$\Delta_T g = \Delta g_p + \Delta g_n$,
$(\Delta g_p + \Delta g_n)/2$, and
$(\Delta g_p + \Delta g_n)/4$, where $p$ and $n$ indicate
proton and neutron respectively,
for calculating the cross section.}
\label{fig:dimuon-dtg-cross}
\vspace{-0.30cm}
\end{figure}

In Figs.\,\ref{fig:dimuon-mass-cross},
\ref{fig:dimuon-pt-cross}, and \ref{fig:dimuon-y-cross},
the gluon transversity in the deuteron is assumed to be
$\Delta_T g = \Delta g_p + \Delta g_n$ for estimating 
the cross sections. However, it is just an assumption.
As mentioned before, the order of magnitude of the quark 
transversity distributions are expected to be similar to
the longitudinally-polarized ones, whereas the gluon transversity
in the deuteron would be very different from the longitudinally
polarized gluon distribution in the nucleon.
In Fig.\,\ref{fig:dimuon-dtg-cross}, we show the cross sections
also by taking $\Delta_T g = (\Delta g_p + \Delta g_n)/2$ 
or $(\Delta g_p + \Delta g_n)/4$. One should note that
the actual cross sections could be very different 
from the estimates by using the assumed gluon transversity 
$\Delta_T g = \Delta g_p + \Delta g_n$.

\begin{figure}[t]
\vspace{-0.00cm}
\begin{center}
   \includegraphics[width=8.0cm]{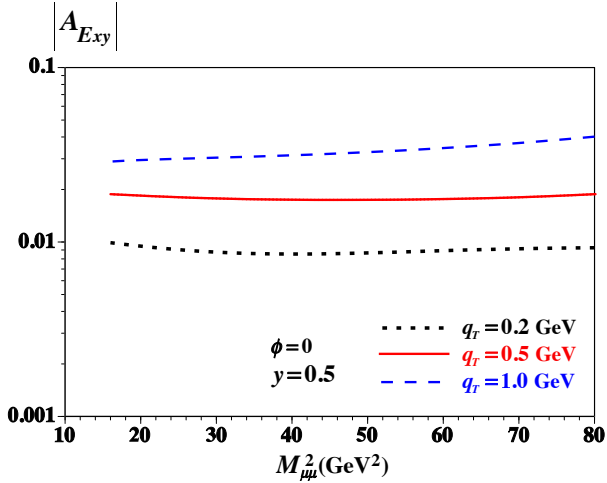}
\end{center}
\vspace{-0.6cm}
\caption{
The spin asymmetries $| A_{E_{xy}} |$ for 
the proton-deuteron Drell-Yan cross sections
are shown as the function of the dimuon-mass squared $M_{\mu\mu}^{\,2}$ at
$q_T=0.2,\, 0.5$, and $1$ GeV for the dimuon azimuthal angle $\phi=0$
and the dimuon rapidity $y=0.5$. The CTEQ14 PDFs are used 
for the unpolarized PDFs of the proton and the deuteron.
The longitudinally-polarized gluon distribution is taken 
from the NNPDF1.1, and the gluon transversity is assumed 
to be the same as the longitudinal one for numerical estimates.}
\label{fig:asym-dy}
\vspace{-0.00cm}
\end{figure}

So far, we have shown the absolute cross sections. However,
actual measurements will be done in polarization asymmetries,
and their numerical results are shown in Fig.\,\ref{fig:asym-dy}
for the polarization asymmetry defined in Eq.\,(\ref{eqn:pol-asym})
as the function of the dimuon-mass squared $M_{\mu\mu}^{\,2}$.
Since the gluon transversity distributions are assumed to be
the same as the longitudinally-polarized ones in this figure,
the numerical values are likely to be most optimistic ones.
The asymmetries are likely to be smaller than these vales.
However, the asymmetries could be within the reach
of future experimental measurements.

The measurement of the gluon transversity is considered at JLab
\cite{jlab-gluon-trans}.
In the studies of nucleon structure functions, the measurements
of hadron-accelerator facilities have been often complementary
and much better in some aspects of the PDFs. 
For example, the pion- and jet-production measurement in polarized
proton-proton collisions at RHIC provided a constraint
on the longitudinally-polarized gluon distribution, namely
the gluon-spin contribution to the nucleon spin. 
Furthermore, the typical $Q^2$ range ($20 < Q^2 < 50$ GeV$^2$) 
of the Drell-Yan experiment is much higher than
the JLab measurements at a few $-$ several GeV$^2$.
In the similar way, it is a good idea to propose independent
experiments to measure the gluon transversity at hadron facilities.
With this motivation, we proposed the Drell-Yan measurement
in the proton-deuteron reaction with the polarized deuteron.
It needs the linearly-polarized deuteron with the unpolarized
proton beam, and the azimuthal-angle information \cite{dy-angular}
is necessary for the dimuon in the final state.
The cross section is typically 0.001 $-$ 0.1 nb/GeV$^2$, so that 
it may not be an easy experiment.
However, there are available hadron facilities at Fermilab, 
J-PARC  (Japan Proton Accelerator Research Complex), 
GSI-FAIR (Gesellschaft f\"ur Schwerionenforschung -Facility for 
Antiproton and Ion Research), and 
NICA (Nuclotron-based Ion Collider fAcility).
In addition, if the fixed-deuteron target becomes possible
at RHIC, Large Hadron Collider (LHC), or EIC, 
there could be a possibility.
We hope that our theoretical proposal is realized 
in future experiments at some facility.

\section{Summary}\label{summary}

Instead of the longitudinally-polarized parton distribution functions,
the nucleon spin structure can be investigated by 
the transversely-polarized ones. 
The leading-twist parton-distribution functions 
in the transversely-polarized nucleon
are quark transversity distributions, which have chiral-odd nature,
and there is some experimental information.
On the other hand, the gluon transversity does not exist for the spin-1/2
nucleon because the two unit of spin flip ($\Delta s=2$) is necessary.
It exists for the spin-1 deuteron. There is an experimental project
to measure the gluon transversity by electron scattering; however,
it is valuable if it can be investigated at hadron-accelerator facilities,
as an independent experiment and to probe the different kinematical region.
There was no theoretical proposal to find it in hadron reactions
before this work.

In this paper, the possibility was proposed to find the gluon transversity
at hadron accelerator facilities, especially, in the proton-deuteron reactions.
We found that it is possible in the proton-deuteron Drell-Yan process 
with the linearly-polarized deuteron. For the final dimuon, 
the experimental measurement on the azimuthal-angle distribution is necessary.
We showed expected dependencies of the cross section on 
the dimuon-mass squared $M_{\mu\mu}^2$, the dimuon transverse-momentum $p_T$,
and the dimuon rapidity $y$, and the assumption on the magnitudes
of the gluon transversity $\Delta_T g$.
The order of cross section was typically estimated as
0.001 $-$ 0.1 nb/GeV$^2$.
Then, the spin asymmetries were shown and they could be within
the reach of experimental measurements.
Hopefully, it will be realized in future experiments
at hadron facilities in addition to the electron scattering
experiment at JLab and possibly at EIC.

\begin{acknowledgements}
The authors thank W.-C. Chang, W. Detmold, D. Keller, P. Mulders,
K. Nakano, J. Qiu, K. Tanaka, W. Vogelsang, and S. Yoshida 
for suggestions.
This work was partially supported by Japan Society for the Promotion 
of Science (JSPS) Grants-in-Aid for Scientific Research (KAKENHI) 
Grant Number 19K03830.
Q.-T.S is supported by the MEXT Scholarship for foreign students 
through the Graduate University for Advanced Studies.
\end{acknowledgements}



\end{document}